\global\def\draftcontrol{0}

   \def\versionno{Lectures on Lambda Positive Universe -- draft   }
\catcode`\@=11

\expandafter\ifx\csname draftcontrol\endcsname\relax\global\def\draftcontrol{0}
\fi

{\count255=\time\divide\count255 by 60
\xdef\hourmin{\number\count255}
\multiply\count255 by-60\advance\count255 by\time
\xdef\hourmin{\hourmin:\ifnum\count255<10 0\fi\the\count255}}
\def\draftdate{\number\month/\number\day/\number\year\ \ \ \hourmin }

\newcommand\makepapertitle{\par
  \begingroup
    \renewcommand\thefootnote{\@fnsymbol\c@footnote}%
    \def\@makefnmark{\rlap{\@textsuperscript{\normalfont\@thefnmark}}}%
    \long\def\@makefntext##1{\parindent 1em\noindent
            \hb@xt@1.8em{%
                \hss\@textsuperscript{\normalfont\@thefnmark}}##1}%
     \newpage
     \global\@topnum\z@   % Prevents figures from going at top of page.
     \@makepapertitle
     \thispagestyle{empty}\@thanks
  \endgroup
  \setcounter{footnote}{0}%
  \global\let\thanks\relax
  \global\let\makepapertitle\relax
  \global\let\@makepapertitle\relax
  \global\let\@thanks\@empty
  \global\let\@author\@empty
  \global\let\@date\@empty
  \global\let\@title\@empty
  \global\let\title\relax
  \global\let\author\relax
  \global\let\date\relax
  \global\let\and\relax
  \def\version{\let\version\@version\@gobble}
}
\def\@makepapertitle{%
  \newpage
   \ifnum\draftcontrol=1 {}
   \version\versionno
   \vskip 3em%
   \else
   \hfill\hbox to 3cm {\parbox{4cm}{\@pubnum}\hss}%
   \vskip 3em%
   \fi
   \begin{center}%
   \let \footnote \thanks
     {\LARGE {\@title}}%
     \vskip 1.5em%
     {\normalsize%\large
       \lineskip .5em%
       \begin{tabular}[t]{l}%
         \@author
       \end{tabular}\par}%
     \vskip 1.5em%
     {\@bstract}%
     \end{center}%
     \vskip 1.5em
     \@date%
   \par
}

\gdef\@pubnum{}
\def\pubnum#1{%
  \gdef\@pubnum{#1}}

\gdef\@bstract{}
\def\Abstract#1{%
  \gdef\@bstract{%
   \parbox{\textwidth-0pc}{%
   \centerline{\bf Abstract}\penalty1000%
\kern.2cm%
\noindent%\abstractfont \baselineskip=12pt
\renewcommand\baselinestretch{1.0}%
{#1}}}
}

\def\ps@paper{\let\@mkboth\@gobbletwo%
     \ifnum\draftcontrol=1
    \def\@oddfoot{\hbox to \textwidth{\tiny \versionno \hfil\tiny\draftdate}%
    \hskip -\textwidth \hbox to \textwidth{\hfil\rm\thepage\hfil}}%
     \else\def\@oddfoot{\hbox to \textwidth{\hfil\rm\thepage\hfil}}
     \fi
     \let\@evenfoot\@oddfoot
}
\def\body{\clearpage
          \pagestyle{paper}
    }
\newenvironment{acknowledgements}{%
\vskip 3.25ex
\noindent {\bf Acknowledgements}
}

\def\@version#1{\ifnum\draftcontrol=1
\typeout{}\typeout{#1}\typeout{}
\vskip3mm\centerline{\hbox{\fbox{\normalsize{\tt DRAFT -- #1 -- }
                   {\draftdate}}}}\vskip3mm
\fi}
\let\version\@version
\long\def\eqlabel#1{\ifnum\draftcontrol=1
                    \tag@false  % there are some problems with multline without this
                    \tag*{(\theequation) \hbox to -0.2cm{\hspace{0cm}\small{#1}\hss}}
                    \refstepcounter{equation}
                    \edef\@currentlabel{\theequation}
                    \ltx@label{#1}          % use old LaTeX \label instead of new definition
                    \else
                    \label{#1}
                    \fi
                    }
\let\st@bibitem\@bibitem
\let\st@lbibitem\@lbibitem
\ifnum\draftcontrol=1
  \def\@bibitem#1{%
    \st@bibitem{#1}\a@@label{#1}\ignorespaces}
  \def\@lbibitem[#1]#2{%
    \st@lbibitem[#1]{#2}\a@@label{#2}\ignorespaces}
  \def\a@@label#1{%
    \gdef\a@lab{\smash{\normalfont\small#1}}
    \ifvmode
      \if@inlabel
        \global\setbox\@labels\hbox{%
          \llap{\a@lab\let\a@lab\relax
                \kern\@totalleftmargin\kern\marginparsep}%
          \box\@labels}%
      \fi
    \fi}
\fi
\documentclass[10pt]{article}
\usepackage{amsmath,amssymb,array,calc,epsfig,bm}
\usepackage{graphicx}        % standard LaTeX graphics tool
\usepackage{multicol}        % used for the two-column index

\usepackage{amsmath,amssymb,bm,epsfig}

\usepackage{cite}
\usepackage[
      colorlinks=true,
      linkcolor=red,  
      urlcolor=blue,    
      filecolor=red,     
      linkcolor=blue,
      citecolor = red,
      pdfstartview=FitV,
      pdftitle={},
        pdfauthor={Paolo Benincasa},
        pdfsubject={},
        pdfkeywords={},
        pdfpagemode=None,
        bookmarksopen=true    
      ]{hyperref}

\usepackage{centernot}      
\usepackage{mathpazo}
\usepackage{stmaryrd}
\usepackage{pifont}
\usepackage{colortbl}
\usepackage{multirow,bigdelim}
\usepackage{arydshln}
\usepackage{tikz, pict2e}
\usepackage{geometry}
\usepackage{lmodern}
\usepackage{wrapfig}
\usepackage{float}
\usepackage{enumitem}
\usepackage[framemethod=default]{mdframed}
\usepackage{array}
\usepackage{mathrsfs}

\usepackage{placeins}
\usepackage{circuitikz}

\usepackage[T1]{fontenc}
\usepackage[utf8]{inputenc}
\usepackage[english]{babel}

\PassOptionsToPackage{table}{xcolor}

\usepackage{xspace}
\usepackage{ifthen}
\usepackage{epstopdf}
\usepackage{amsthm}

\usepackage{relsize}
\usepackage{tabularx}
\usepackage{booktabs}
\usepackage{soul}
\usepackage{pdfpages}
\usepackage[most]{tcolorbox}
\usepackage{capt-of}
\usepackage{xfrac}
\usepackage{empheq}
\usepackage{slashed}
\usepackage{cancel}
\usepackage{upgreek}
\usepackage[skip=1em]{parskip}

\usepackage{latexsym}
\usepackage{ifoddpage}
\usepackage{subcaption}

\usepackage{cancel}
\usepackage{comment}
\usepackage{shuffle}

\usepackage{savesym}
\savesymbol{Ref}
\savesymbol{erf}
\usepackage{physics}
\restoresymbol{phys}{Ref}
\restoresymbol{phys}{erf}

\usepackage[compat=1.1.0]{tikz-feynman}
\tikzfeynmanset{compat=1.1.0}

\definecolor{myblue}{RGB}{166, 193, 222}
\definecolor{myblue2}{RGB}{68, 114, 196}
\definecolor{mygreen}{RGB}{175, 218, 162}
\definecolor{mygreen2}{RGB}{133, 210, 110}
\definecolor{myorange2}{RGB}{255, 192, 0}

\newtcolorbox{shade}[1][]{
  top=-7pt,bottom=6pt,
  sharp corners=all,
  colback=white!95!black,
  colframe=white,
}

\newcolumntype{P}[1]{>{\centering\arraybackslash}p{#1}}
\newcolumntype{L}{>{$}l<{$}}

\usetikzlibrary{calc,backgrounds}
\usetikzlibrary{arrows,shapes,positioning,shadows,trees}
\usetikzlibrary{arrows.meta, shapes.geometric}
\usetikzlibrary{mindmap}
\usetikzlibrary{decorations.pathreplacing}
\usetikzlibrary{decorations.pathmorphing}
\usetikzlibrary{decorations.markings}
\usetikzlibrary{matrix}
\usetikzlibrary{patterns.meta}

\tikzstyle arrowstyle=[scale=1]
\tikzstyle directed=[postaction={decorate,decoration={markings,
    mark=at position .5 with {\arrow[arrowstyle]{stealth}}}}]
\tikzstyle reverse directed=[postaction={decorate,decoration={markings,
    mark=at position .5 with {\arrowreversed[arrowstyle]{stealth};}}}]

\ifnum\draftcontrol=1
\fi
\renewcommand\baselinestretch{1.25}
\renewcommand\section{\@startsection {section}{1}{\z@}%
                                   {-3.5ex \@plus -1ex \@minus -.2ex}%
                                   {2.3ex \@plus.2ex}%
                                   {\normalfont\large\bfseries}}
\renewcommand\subsection{\@startsection{subsection}{2}{\z@}%
                                   {-3.25ex\@plus -1ex \@minus -.2ex}%
                                   {1.5ex \@plus .2ex}%
                                   {\normalfont\normalsize\bfseries}}
\renewcommand\subsubsection{\@startsection{subsubsection}{3}{\z@}%
                                   {-3.25ex\@plus -1ex \@minus -.2ex}%
                                   {1.5ex \@plus .2ex}%
                                   {\normalfont\normalsize\it}}
\renewcommand\paragraph{\@startsection{paragraph}{4}{\z@}%
                                   {-3.25ex\@plus -1ex \@minus -.2ex}%
                                   {1.5ex \@plus .2ex}%
                                   {\normalfont\normalsize\bf}}

\numberwithin{equation}{section}

\def\revise#1       {\raisebox{-0em}{\rule{3pt}{1em}}%
                     \marginpar{\raisebox{.5em}{\vrule width3pt\
                     \vrule width0pt height 0pt depth0.5em
                     \hbox to 0cm{\hspace{0cm}{%
                     \parbox[t]{4em}{\raggedright\footnotesize{#1}}}\hss}}}}

\def\sqr#1#2{{\vcenter{\vbox{\hrule height.#2pt
 \hbox{\vrule width.#2pt height#1pt \kern#1pt
 \vrule width.#2pt}\hrule height.#2pt}}}}

\catcode`\@=12

\newcommand{\ee}{e}

\newcommand{\beq}{\begin{equation}}
\newcommand{\eeq}{\end{equation}}
\newcommand{\bea}{\begin{equation}\begin{aligned}}
\newcommand{\eea}{\end{aligned}\end{equation}}

\newlength{\wsingfig}
\newlength{\wdblefig}
\newlength{\wquadfig}
\newlength{\wtriplefig}
\newcommand   \lptl{\raise .8ex\hbox{$^\leftarrow$} \hspace{-9pt} \partial}
\newcommand   \lrptl{\raise .8ex\hbox{$^\leftrightarrow$} \hspace{-9pt} \partial}

\newcommand   \be {\mathbf{e}}

\def\be#1\ee{\begin{align}#1\end{align}}
\newcommand \nn {\nonumber}

\newcounter{ssooeexercise}[section]
\renewcommand{\thessooeexercise}{\thesection.\arabic{ssooeexercise}}

\newtcolorbox{exbox}{
  colback=gray!10,
  colframe=gray!20,
  boxrule=1pt,
  arc=5pt,
  breakable,
  left=1em,
  right=1em,
  top=0.8em,
  bottom=0.8em
}

\newcommand{\newexercise}[3][]{%
  \refstepcounter{ssooeexercise}%
  \expandafter\gdef\csname ex@content@#2\endcsname{#3}%
  \expandafter\xdef\csname ex@number@#2\endcsname{\thessooeexercise}%
  \begin{exbox}%
    \textbf{Exercise~\thessooeexercise:} #3
    \ifx&#1&\else\label{#1}\fi
  \end{exbox}%
}

\newcommand{\printexercise}[1]{%
  \begin{exbox}%
    \textbf{Exercise~\csname ex@number@#1\endcsname:}
    \csname ex@content@#1\endcsname
  \end{exbox}%
}

\begin{document}

%%%%%%%%%
%%%%%% text starts here
%%%%%%%%%

\title{\underline{\Large\bf The Disordered Universe 2025:} \\
       {\Large\bf Features in a $\Lambda>0$ Universe}}

\pubnum{%
}

\author{%
	{\scshape Lecturer:} {\bf Dionysios Anninos}${}^{\circledast}$,%
    \\
    {\scshape Notes written by:} {\bf Samuel Brian}${}^{\circledast}$, {\bf Yakov Landau}${}^{\star}$, {\bf Tommaso Marini}${}^{\ast}$, \\
      \hspace{3.375cm} {\bf Daniel McLoughlin}${}^{\odot}$, {\bf Polina Petriakova}${}^{\oplus}$, {\bf Rajat Sharma}${}^{\diamond}$
    \\
    {\scshape Edited by:} {\bf Paolo Benincasa}${}^{\dagger\,\ddagger}$, {\bf Guilherme Leite Pimentel}${}^{\diamond}$
    \\ [0.4cm]
    \ttfamily ${}^{\circledast}$ Department of Mathematics, \\
    \ttfamily $\phantom{{}^{\dagger}}$ King’s College London, Strand, London WC2R 2LS, UK\\
    \ttfamily ${}^{\star}$ Walter Burke Institute for Theoretical Physics, \\
    \ttfamily $\phantom{{}^{\dagger}}$ Caltech, Pasadena, California 91125, USA \\
    \ttfamily ${}^{\ast}$ Department of Physics, Princeton University,\\
    \ttfamily $\phantom{{}^{\dagger}}$ Princeton, NJ 08544, USA  \\
    \ttfamily ${}^{\odot}$ Center for Theoretical Physics, Department of Physics,\\
    \ttfamily $\phantom{{}^{\dagger}}$ Columbia University, Pupin Hall, 538 West 120th Street, New York, NY 10027, USA\\ 
    \ttfamily ${}^{\oplus}$ Bogoliubov Laboratory of Theoretical Physics, Joint Institute for Nuclear Research, \\
    \ttfamily $\phantom{{}^{\dagger}}$ Joliot-Curie Str. 6, 141980 Dubna, Russian Federation \\
    \ttfamily ${}^{\dagger}$ Asociaci{\'o}n Cultural La Ricotta, Espazo La Ricotta\\
    \ttfamily $\phantom{{}^{\dagger}}$ As Barreiras, Castro Caldelas, E-32769 Galicia-Spain\\
    \ttfamily ${}^{\ddagger}$ Instituto Galego  de F{\'i}sica de Altas Enerx{\'i}as IGFAE, \\
    \ttfamily $\phantom{{}^{\dagger}}$ Universidade de Santiago de Compostela, E-15782 Galicia-Spain\\
    \ttfamily ${}^{\diamond}$ Scuola Normale Superiore, \\
    \ttfamily $\phantom{{}^{\dagger}}$ Piazza dei Cavalieri 7, 56126, Pisa, Italy\\[0.4cm]
    \small \ttfamily dionysios.anninos@kcl.ac.uk, samuel.brian@kcl.ac.uk, ylandau@caltech.edu, tm8158@princeton.edu,\\
    \small \ttfamily dcm2183@columbia.edu, polinapetriakova@gmail.com, rajat.sharma@sns.it, \\
    \small \ttfamily pablowellinhouse@sindominio.net, guilherme.leitepimentel@sns.it \\ [0.75cm]
}

\Abstract{%
A complete understanding of cosmology, from its earliest epoch all the way to the current dark energy dominated era, requires the development of a mathematical framework that incorporates quantum features in an expanding universe. Such a framework, alongside a description of the underlying degrees of freedom, is so far lacking. Several of the pertinent challenges and puzzles already arise upon considering even the most idealised model of an expanding universe --- the de Sitter spacetime and fluctuations thereof. In these lectures, we review some elementary ideas of the de Sitter physics, passing through the representation theory of the de Sitter group, aspects of canonical quantum gravity, all the way to attempts in describing {quasi-local} features. Throughout, we highlight open problems that motivate ongoing efforts to understand quantum gravity and cosmology in universes with $\Lambda>0$. These notes are based on a series of lectures held during the \href{https://indico.sns.it/event/91/}{{\it La Ricotta} Summer School: {\it The Disordered Universe 2025}} from 21~July to 1~August~2025.
}

\makepapertitle

\body

\version\versionno

\newpage

\section*{Preface}
This article is a chapter from the book {\it The Disordered Universe: Open Systems, Scattering Amplitudes, de Sitter Space, and Disordered Systems in Cosmology}~\cite{Benincasa:2027tdu}, which collects the lectures delivered at the \href{https://indico.sns.it/event/91/}{\it La Ricotta Summer School: The Disordered Universe 2025}, held from 21~July to 1~August~2025 at the \href{https://blogs.sindominio.net/laricotta/#}{\it Espazo La Ricotta} in As Barreiras (Castro Caldelas, Spain). These lecture notes are intended as a pedagogical introduction to topics that are becoming increasingly relevant in cosmology. 

\newpage

\tableofcontents

\section{Cosmological origins}

The motivation behind our topic dates back over a century. In 1922, Friedmann demonstrated that under the assumptions of spatial homogeneity and isotropy on sufficiently large scales, the field equations of Einstein's general relativity in the presence of a cosmological constant\footnote{Throughout these lecture notes we will set $c=\hbar=1$, and take repeated indices to be summed over unless otherwise specified. By $G$ we denote the Newton constant, and by $\Lambda$ the cosmological constant. When working in general spacetime dimensions, we label their total number by $d$.}
\begin{equation}\label{Einstein_equations}
R_{\mu\nu} - \frac{1}{2} R  g_{\mu\nu} + \Lambda  g_{\mu\nu} = 8 \pi G   T_{\mu\nu}~,  
\end{equation} %naturally 
admit non-stationary solutions of the following type
\begin{equation}\label{FLRW_metric}
ds^2 = - dt^2 + a(t)^2 d \Sigma^2_3~.
\end{equation}
Here, $d \Sigma^2_3$ is a metric of a three-dimensional space of constant curvature, which could be a three-sphere $\mathbb{S}^3$, a flat Euclidean space $\mathbb{R}^3$, or a hyperbolic space $\mathbb{H}^3$.\footnote{A little more specifically, Friedmann investigated the case of positive spatial curvature in 1922 \cite{Friedmann:1922}, and negative spatial curvature in 1924 \cite{Friedmann:1924}. A positively curved spatial slicing was also considered in 1927 by  Lema\^{i}tre \cite{lemaitre1927univers}. The possibility of a flat slicing appeared in 1931 \cite{heckmann1931metrik}. Around the mid 1930's Robertson \cite{robertson1935kinematics} and Walker \cite{walker1937milne} established that the metric (\ref{FLRW_metric}) is uniquely fixed under the assumption of spatial homogeneity and isotropy.} The scale factor, $a(t)$, is then found to obey a simple ordinary differential equation that we will momentarily discuss. The theoretical framework for an expanding spacetime was independently rediscovered in 1927 by Lema\^{i}tre~\cite{lemaitre1927univers}, who was the first to derive the linear redshift–distance relation, transforming the abstract concept of a dynamical scale factor $a(t)$ into a concrete testable prediction. Remarkably, around almost the same time, Hubble’s observations~\cite{Hubble:1929} of distant galaxies provided the first empirical evidence for cosmic expansion, confirming that the large-scale structure of the universe indeed exhibits considerable time dependence. Ever since the 1920's, all subsequent observations of distant  bodies have consistently supported the picture that we inhabit an expanding Universe. By demonstrating that Einstein's theory provides a mathematical basis for an expanding spacetime, the theory of cosmology begins its course.

To determine a scale factor $a(t)$ in Friedmann's ansatz~\eqref{FLRW_metric}, one must specify the type of matter fields in Einstein's field equations~\eqref{Einstein_equations} and their respective equation of state. In the presence of a matter stress-energy tensor that takes the form of a perfect fluid in the comoving frame, one has $T_{\mu\nu} = \text{diag}\bigl(\rho(t), p(t), p(t), p(t)\bigr)$. Here $\rho(t)$ represents the energy density, and $p(t)$ represents the pressure of the fluid in the comoving frame, and they are assumed to be purely time dependent in equilibrium.  The temporal component of~\eqref{Einstein_equations} is
\begin{equation}\label{Friedmann_eq_1}
-\, \biggl(\frac{\dot{a}}{a}\biggr)^2 - \frac{k}{a^2}  + \frac{8\pi G }{3} \rho + \frac{\Lambda}{3}= 0~.
\end{equation} 
Equation (\ref{Friedmann_eq_1}) is known as the Friedmann equation, and the parameter $k= \{ +1, 0 , -1 \}$ labels the corresponding 
curvature sign of the spatial three-metric in~\eqref{FLRW_metric}. One way to interpret (\ref{Friedmann_eq_1}) is that in general relativity the bulk energy vanishes due to time-reparametrisation invariance. The first term with the time-derivative is a negative kinetic term from the metric, while the remaining terms are akin to a potential energy.   

The spatial components of~\eqref{Einstein_equations} are not independent of~\eqref{Friedmann_eq_1}. The conservation law for the stress-energy tensor, $\nabla^\mu T_{\mu\nu} = 0$, often referred to as the continuity equation, leads to the second equation
\begin{equation}
    \dot{\rho} +3 \, \dfrac{\dot{a}}{a} \, \bigl(\rho+p\bigr) = 0~. 
\end{equation} For an equation of state of the type $p = w \rho$, which is true for various models of interest,  the continuity equation requires the energy density to evolve as $\rho \propto a^{-3(1+w)}$. Particularly, $w = \tfrac{1}{3}$ corresponds to a radiation dominated Universe, $w = 0$ to a Universe dominated by pressureless dust, and $w = -1$ to one dominated by the cosmological constant, $\Lambda$. Upon implementing the equation of state, one subsequently computes the scale factor $a(t)$ from Friedmann's equation~\eqref{Friedmann_eq_1}, which thus encodes substantial physics.
%\begin{equation}   a_{\text{R}}(t) \propto t^{1/2}, \qquad a_{\text{D}}(t) \propto t^{2/3}, \qquad a_\Lambda(t) \propto .\end{equation}

From this scaling behaviour, we learn that at sufficiently large values of $a(t)$, the contribution of the $\Lambda${--}term will dominate. To see this, we note that the energy density has different components, and evolves as
\begin{equation} \rho \, %= \, \frac{3 H_0^2}{8 \pi G} 
\propto \, %\biggl( 
\Omega_{\Lambda} + \Omega_{\text{curv}} \, \Bigl( \frac{a_0}{a}\Bigr)^2 + \Omega_{ \text{dust}} \Bigl( \frac{a_0}{a}\Bigr)^3 + \Omega_{\text{rad}} \Bigl( \frac{a_0}{a}\Bigr)^4 %\biggr)
, \end{equation}
where $a_0=a(t_0)$ is the value at some initial time. The dimensionless parameters $\Omega_i$ represent the fractional contribution of each component to the total energy density budget. Consequently, there exists a source of energy, known as the cosmological constant $\Lambda$, potentially realised in nature \cite{SupernovaSearchTeam:1998fmf,SupernovaCosmologyProject:1998vns,Planck:2018vyg-ch04}, that dominates the behaviour of a cosmological spacetime at late times. The reason is that, when constant, $\Lambda$ does not dilute, and as such it yields an attractor late-time behaviour for cosmological dynamics. Therefore, under fairly general conditions, a universe with a non-zero cosmological constant will eventually evolve into a %de Sitter-like 
state dominated by this persistent component. Current observational data is currently consistent with this paradigm. Further evidence, however, is subject to more precise statistics and/or future data coming from ongoing experiments.

From a Lagrangian formulation of general relativity, the cosmological constant term arises in a rather simple and natural way. Viewed as a low energy effective field theory, general relativity coupled to matter fields is a leading term in an infinite series of increasingly irrelevant higher-derivative terms,
\begin{equation}
S_{\text{EFT}} = \frac{1}{16\pi G} \int d^4 x \, \sqrt{-g} \, \Bigl( - 2 \Lambda + R + \ldots \Bigr) + S_{\text{matter}}~.
\end{equation} 
The higher derivative terms have dimensionful coefficients dictating at what scales they begin to play an important role. Within any effective theory of gravity coupled to matter, the $\Lambda$-term corresponds to a coupling to the covariant volume element. The $\Lambda$-term respects all symmetries, it is gauge-invariant, and constitutes the leading relevant operator in the effective field theory.\footnote{If there are scalar fields $\varphi$, in the theory, one has a potential competition to $\Lambda$ depending on the features of the scalar potential $V(\varphi)$. Here we are assuming that such a $V(\varphi)$ has a minimum about which the classical theory resides. Whether or not this is a global minimum, and the implications thereof is an interesting question. A physically relevant case is the nature of the Higgs potential. Also, in the inflationary scenario one often postulates the presence of a slowly varying scalar potential of the inflaton field.} The $\Lambda$--term inevitably dictates the dynamics on the largest cosmological scales, and there exists no known symmetry or physical principle that would naturally force us to discard it. 

The $\Lambda$-term has had a long and tumultuous history since its inception. In 1916, Nernst already imagined the effect of a gravitating quantum mechanical vacuum energy \cite{nernst1916versuch}. In the hopes of creating a static and closed universe, in 1917 Einstein modified his field equations to contain a $\Lambda$-term \cite{einstein1917kosmologische}. In 1917, de Sitter noted that Einstein's modified equations admit a static world that can be supported without matter, albeit one with gravitational redshift effects \cite{deSitter:1917}. The de Sitter solution was independently discovered by Levi-Civita in 1917 \cite{levi1917realta}. It was eventually understood, by the mid 1920's, that de Sitter's solution describes an expanding spacetime by Lanczos \cite{lanczos1924stationare} and Lema\^itre \cite{lemaitre1925note}, among others. In 1926, Lens noted that, in order to be observationally viable, $\Lambda$ would need a peculiarly small value  \cite{lenz1926gleichgewicht}. By the early 1930's, following Hubble's measurements of an expanding Universe, Einstein abandoned the cosmological term. By the 1960's, astronomical observations led people to  investigate the cosmological constant, and to the idea that it receives contributions from non-vanishing vacuum energy density~\cite{1967ApJ...147.1222P}. Zel'dovich proposed to identify this non-zero vacuum energy density with fluctuations of quantum fields~\cite{Zeldovich:1967,Zeldovich:1968}. This vacuum energy gravitates, and its contribution to the stress-energy tensor is formally equivalent to the cosmological constant term in Einstein's field equations (\ref{Einstein_equations}).\footnote{Unbeknownst to Einstein, an analogous modification of the Newton's gravitational equation had been discussed by Neumann \cite{neumann1896allgemeine} and Seelinger \cite{seeliger1895newton} in 1896.} According to this view, the cosmological constant $\Lambda$ is no longer a free geometrical parameter but a physical quantity, to be estimated from the quantum field theory. In 1984, Hawking 'proved' that the cosmological constant is vanishing \cite{hawking1984cosmological}, and this proof was disputed in \cite{duff1989cosmological} though the article maintained that $\Lambda$ probably vanishes. In the late 1980's, Weinberg offered an anthropic perspective that anticipated a non-zero, positive value of the cosmological constant~\cite{Weinberg:1989}. 

In the late 1990's, the value of $\Lambda \approx 10^{-52}\text{m}^{-2} \approx 10^{-122} \, \ell^{-2}_{\text{Pl}}$, with $\ell_{\text{Pl}}$ being the Planck length, was deduced from supernova distance-redshift data, establishing the current phase of accelerated expansion  of our Universe~\cite{1998ApJ...507...46S,Riess:1998,Perlmutter:1999}. The $\Lambda$-term received further support from the CMB observations of the WMAP \cite{bennett2013nine} and Planck \cite{Planck:2018vyg-ch04} experiments, as well as the baryon acoustic oscillations \cite{eisenstein2005detection}.  It is an ongoing experimental effort to confirm, or refute, that $\Lambda$ is indeed constant. The measured value of $\Lambda$ in Planck units is strikingly small, completely insensitive to any ultraviolet energy scale that invalidates our current low energy effective field theory. Aside from anthropic reasoning, there is no obvious and universally agreed upon cancellation mechanism that naturally leads to such a small value for $\Lambda$. The quest for a mechanism that explains the measured value of $\Lambda$ is often referred to as the cosmological constant problem.

Another phase of accelerated expansion occurs during the primordial cosmological era of the Universe, according to the inflationary scenario  \cite{Guth:1980zm-ch04,Starobinsky:1980te-ch04,Linde:1981mu-ch04,Albrecht:1982wi-ch04} (for a pedagogical overview see~\cite{Baumann:2009}). In inflationary cosmology,  classical features of the current Universe have a quantum mechanical origin, and inflation offers a mechanism for generating the primordial density perturbations that lead to the formation of the large-scale structure~\cite{Mukhanov:1981,Guth:1982ec-ch04,Hawking:1982cz-ch04,Bardeen:1983qw-ch04}.\footnote{Pioneering pre-inflation ideas, dating back to the 1970's, on primordial cosmological fluctuations date back to the work of Harrison \cite{harrison1970fluctuations}, Zel'dovich \cite{zeldovich1972hypothesis}, and Peebles and Yu \cite{Peebles:1970ag}. They hypothesised that the fluctuations were exactly scale invariant, whereas the observed spectrum is tilted slightly away from a perfectly scale invariant spectrum.} Current experimental evidence provides a lower bound on the duration of inflation, commonly expressed as a requirement of $n_e \gtrsim 50$ e-folds by the end of inflation. The absence of a corresponding upper bound presents another cosmological challenge: the inflationary scenario allows for a potentially vast, unobservable deep infrared phase of the early universe, whose physical details remain inaccessible \cite{guth2007eternal}. Moreover, the fluctuations of inflation also include those of the spin-two linearised metric. Although we have not observed these tensor modes in the cosmic microwave background (CMB), they are part of the inflationary scenario,  and are actively being searched for. As such, in addition to the quantised inflaton field, a proper understanding of inflation requires an understanding of the quantised metric field. It is intimately tied to the problem of quantum gravity.

To date, there is no complete mathematical formulation of cosmology at the quantum level. We do not have an answer, or a comprehensive toolkit, to address the question: {\it What is cosmology?} This defines a central challenge in modern theoretical physics. These lecture notes are motivated by this challenge, and aim to discuss some foundational concepts.  To capture quantum features of an expanding universe in a sharp way, in what follows we will take a theoretical route and explore an idealised version of a purely $\Lambda${--}driven cosmology. This is a good approximation for both the inflationary and the current accelerated phases. It is also the most mathematically tractable case. We shall start with the classical features, a rich subject in its own right, before proceeding to examine quantum field theoretic properties. 
\newline\newline
{\textbf{Lectures.}} These lectures have drawn from a large literature, and we will try to include key references as we proceed. Moreover, there are several modern reviews that we have also drawn from including \cite{Spradlin:2001pw,Anninos:2012qw,Galante:2023uyf,Baumann:2022jpr-ch04}, as well as the comprehensive treatment in \cite{beatrix}. The lectures were delivered in an informal fashion that paralleled the inspiring philosophy of the \href{https://indico.sns.it/event/91/}{Disordered Universe school} itself. It is an accomplishment of the students, and entirely to their credit, that they collected detailed enough notes to reconstruct these lectures in both spirit and form. The topics discussed are mostly foundational, building the necessary background to appreciate modern ideas which include: if and how the holographic principle applies to cosmology, the Gibbons-Hawking proposal \cite{Gibbons:1977mu} for the de Sitter event horizon entropy, modern quantum field theoretic tools, and novel tractable models (see \cite{Anninos:2017hhn,Coleman_2022,Narovlansky:2023lfz,Susskind:2022bia,Collier:2025lux} for a recent sample). To explore these modern topics, we strongly encourage the reader to consult \cite{Baumann:2022jpr-ch04,Galante:2023uyf,beatrix} and the references therein. 
%There are several open problems. %quantum features of gravity, QFT in dS,

\section{de Sitter geometry}\label{sec:pure_dS}
%%= \frac{d-1}{l^2} \, g_{\mu\nu}$. 
The maximally symmetric vacuum solution to Einstein's field equations~\eqref{Einstein_equations} with a positive cosmological constant, $\Lambda >0$, is de Sitter space. In this section, we describe its classical geometry. In $d>2$ spacetime dimensions, taking the trace of equations~\eqref{Einstein_equations} yields $R = \tfrac{2 d}{(d-2)} \, \Lambda$. Since the de Sitter solution has a constant and positive Ricci scalar, we can rewrite \eqref{Einstein_equations} as $R_{\mu\nu} = \frac{2 \Lambda}{(d-2)} \, g_{\mu\nu}$. Spacetime metrics obeying this equation are known as Einstein metrics. It is also useful to define a de Sitter radius, $\ell$, that is related to the cosmological constant $\Lambda$ as
\begin{equation} \Lambda \equiv \frac{(d-2)(d-1)}{2\ell^2}~. \end{equation} 
The de Sitter spacetime can be realised as the induced metric on a $d$-dimensional hypersurface embedded in a $(d+1)$-dimensional Minkowski space. The embedding equation governing the hypersurface is  given by
\begin{equation}\label{hypersurface_eq}
-  \bigl(x^0\bigr)^2 +\delta_{ij} x^i  x^j = \ell^2~,  \quad\quad i = 1,\ldots,d~.
%\, \sum_{i=1}^{d} \, \bigl(x^i\bigr)^2 
\end{equation}
The $(d+1)$-dimensional ambient Minkowski space, $\mathbb{R}^{1,d}$ is described by the metric
\begin{equation}\label{interval_flat}
ds^2 = - \, \bigl(dx^0\bigr)^2 +\delta_{ij}\, dx^i d x^j~.
\end{equation}
%Different insights into the structure of de Sitter space are revealed through its various coordinate representations. 
Below, we  list a few of the coordinate systems that cover portions of the de Sitter space for the case $d=4$. 

\subsection{Global slicing}
A convenient parametrisation of~the embedding coordinates is
\begin{align}\label{global_coord_param}
    x^0 &= \ell \, \sinh{\frac{\tau}{\ell}}~, \nonumber \\
    x^1 &= \ell \, \cosh{\frac{\tau}{\ell}}
    \sin{\theta_1} \sin{\theta_2}\cos{\theta_3}~, \nonumber \\
    x^2 &= \ell \, \cosh{\frac{\tau}{\ell}} \sin{\theta_1} \sin{\theta_2}\sin{\theta_3}~, \\ \nonumber 
    x^3 &= \ell \, \cosh{\frac{\tau}{\ell}}\sin{\theta_1} \cos{\theta_2}~, \\ \nonumber 
    x^4 &= \ell \, \cosh{\frac{\tau}{\ell}}\cos{\theta_1}~,
\end{align} 
where $\tau \in\mathbb{R}$ is called global time, while $\theta_i$ are the angular parameters on a unit sphere defined as $0 \leq \theta_{i}\leq \pi $ for $1\leq i\leq(d-1)$ and $ 0 \leq \theta_{d}<2\pi$. One can check that the chosen coordinates satisfy~\eqref{hypersurface_eq}. 
Plugging~\eqref{global_coord_param} into~\eqref{interval_flat}, we obtain the induced metric on the hyperboloid:
\begin{equation}\label{global_dS_metric}
ds^2 = - d\tau^2 + \ell^2\cosh^2{\frac{\tau}{\ell}}  d\Omega_3^2~, 
\end{equation} where $d\Omega_3^2 = d\theta_1^{\,2} + \sin^2{\theta_1} \, d\theta_2^{\,2} + \sin^2{\theta_1} \sin^2{\theta_2} \, d\theta_3^{\,2}$ is the metric on a unit three-sphere~$\mathbb{S}^{3}$. Consequently, the topology of this slicing is $\mathbb{R} \times \mathbb{S}^3$, and these coordinates cover the entire hyperboloid. On constant time slices, the spatial geometry is a three-dimensional sphere of radius $\ell \cosh{\frac{\tau}{\ell}}$. This sphere contracts from infinite size at $\tau = -\infty $, reaches a minimum radius $\ell$  at $\tau = 0$, and then expands again to infinite size as $\tau \rightarrow +\infty$.
Metric~\eqref{global_dS_metric} therefore corresponds to a bouncing closed Friedmann universe~\eqref{FLRW_metric} with $k=+1$ and the corresponding expansion law. It is the geodesically complete form of de Sitter space.

\subsection{Flat slicing}
Another useful coordinate choice is the planar coordinates or the flat slicing, given by the parametrisation:
\begin{align}\label{flat_coord_param}
x^0 &= \ell \, \sinh{\frac{t}{\ell}} + \frac{{e}^{\frac{t}{\ell}}}{2\ell} \, \delta_{ij} y^i y^j~ , %\Bigl((y^1)^2 + (y^2)^2 + (y^3)^2 \Bigr)~,
\nonumber \\
x^i &= y^i \, {e}^{\, \frac{t}{\ell}}, \qquad \, i={1,2,3}~, \\ \nonumber 
x^4 &=  \ell \, \cosh{\frac{t}{\ell}} - \frac{{e}^{\frac{t}{\ell}}}{2\ell} \, \delta_{ij} y^i y^j~. %\Bigl((y^1)^2 + (y^2)^2 + (y^3)^2 \Bigr)~.
\end{align} 
The time coordinate $t\in\mathbb{R}$, and the spatial coordinates $y^i$ cover $\mathbb{R}^3$. This leads to the metric of a flat Friedmann universe with exponential expansion:
\begin{equation}\label{flat_dS_metric}
    ds^2 = - dt^2 + {e}^{\frac{2t}{\ell}}  \delta_{ij} dy^i dy^j~.
\end{equation} Planar coordinates~\eqref{flat_coord_param} cover only half of the hyperboloid, since $x^0+x^4 =\ell  {e}^{\frac{t}{\ell}}>0$. On a constant time slice, the spatial geometry is an infinite, flat 3‑plane, $\mathbb{R}^3$. Note that in the limit $\ell \rightarrow \infty$ at fixed $t$, we find ourselves in the Minkowski spacetime. For completeness, we also note that using the conformal time $\eta= -\ell{e}^{ - \frac{t}{\ell}}$, $\eta \in \mathbb{R}^-$, metric~\eqref{flat_dS_metric} becomes
\begin{equation}\label{flat_dS_metric_conformal_time}
    ds^2= \frac{\ell^2}{\eta^2} \, \bigl( - \, d\eta^2 + \delta_{ij} dy^i dy^j\bigr)~.
\end{equation} %here $- \infty < \eta < 0$. 
The planar metric corresponds to a flat hyperbolic Friedmann universe~\eqref{FLRW_metric} with $k=0$. 

\subsection{Hyperbolic slicing}
Introducing the hyperbolic coordinates, known as the open or Milne-type slicing\footnote{The Milne universe is a solution to the vacuum Einstein’s field equations, with the metric \begin{equation}\label{Milne_metric}
ds^2 = - dt^2 + t^2 dH^2_3~,  
\end{equation} which is the hyperbolic slicing of the Minkowski spacetime.}
\begin{align}\label{hyperbolic_coord_param}
    x^0 &= \ell \, \sinh{\frac{\tilde{\tau}}{ \ell}} \cosh{\tilde{\theta}_1}~, \nonumber \\
    x^1 &=  \ell \, \sinh{\frac{\tilde{\tau}}{ \ell}} \sinh{\tilde{\theta}_1} \sin{\theta_2}\cos{\theta_3}~, \nonumber \\
    x^2 &=  \ell \, \sinh{\frac{\tilde{\tau}}{ \ell}} \sinh{\tilde{\theta}_1} \sin{\theta_2}\sin{\theta_3}~, \\ \nonumber 
    x^3 &=  \ell \, \sinh{\frac{\tilde{\tau}}{ \ell}} \sinh{\tilde{\theta}_1} \cos{\theta_2}~, \\ \nonumber 
    x^4 &=  \ell \, \cosh{\frac{\tilde{\tau}}{ \ell}}~,
\end{align} 
with $\tilde{\tau} \in \mathbb{R}^+$ and $\tilde{\theta}_1 \in \mathbb{R}^+$, while $\theta_2$ and $\theta_3$ are angular coordinates on $S^2$. We obtain the induced metric 
\begin{equation}\label{hyperbolic_dS_metric}
ds^2= - \, d\tilde{\tau}^2 +  \ell^2 \sinh^2{\frac{\tilde{\tau}}{ \ell}} \, d H_3^2~.
\end{equation} Here $d H_3^2 = d\tilde{\theta}_1^{\,2} + \sinh^2{\tilde{\theta}_1} \left( d\theta_2^{\,2} +  \sin^2{\theta_2} \, d\theta_3^{\,2} \right)$ is the metric on a unit three-dimensional hyperboloid $\mathbb{H}^3$, with $0 \leq \tilde{\theta}_1< \infty$ and the usual angular ranges $0 \leq \theta_2\leq \pi $ and $ 0 \leq \theta_3 < 2\pi$. The  coordinates~\eqref{hyperbolic_coord_param} cover only half of the hyperboloid since $x^0>0$ and $x^4>0$. The obtained metric corresponds to an open hyperbolic Friedmann universe~\eqref{FLRW_metric} with $k=-1$. 

We emphasise that two of these three homogeneous and isotropic slices, namely,~\eqref{flat_dS_metric} and~\eqref{hyperbolic_dS_metric}, have a monotonic scale factor, while~\eqref{global_dS_metric} has a bounce.
\newline\newline
{\textbf{de Sitter isometries, briefly.}} From the embedding space perspective, four-dimensional de Sitter space is defined as a hypersurface in $\mathbb{R}^{1,4}$. The symmetry group preserving this hypersurface equation~\eqref{hypersurface_eq} is the $SO(1,4)$ Lorentz subgroup of the five-dimensional Poincar\'e group. Consequently, the isometry group of four-dimensional de Sitter space is $SO(1,4)$, whose ten generators make it a maximally symmetric space. 
%As expected for a $d$-dimensional maximally symmetric space, these ten Killing vectors decompose into $d(d-1)/2$ generators of the isotropy subgroup and $d$ generators that act analogously to translations. 
The various coordinate systems of four-dimensional de Sitter space make different subgroups of its full $SO(1,4)$ isometry manifest. The global coordinates~\eqref{global_coord_param} with its $\mathbb{S}^3$ foliation (see metric~\eqref{global_dS_metric}) manifest the maximally compact rotational subgroup $SO(4)$. In~contrast, the flat slicing~\eqref{flat_dS_metric} manifests a different subgroup, the six $ISO(3)$ symmetries of the three-dimensional Euclidean space, combined with a dilatation symmetry $SO(1,1)$. Similarly, the hyperbolic coordinates~\eqref{hyperbolic_coord_param} manifest yet another subgroup, $SO(1,3)$. While each coordinate systems are useful for revealing specific subgroups of the $SO(1,4)$ de Sitter group, none of them makes the full ten-parameter isometry group manifest.

\subsection{Causal structure \& the static patch}
\begin{figure}[t] \centering
	\centering
	\includegraphics{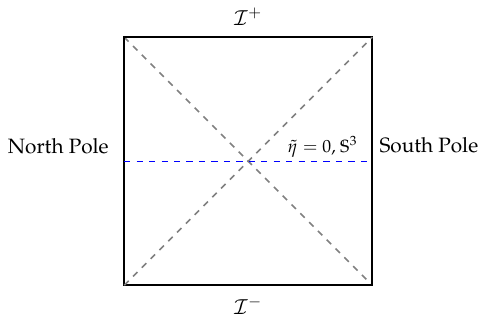}
\caption{The blue line (like all horizontal lines) depicts a slice of $\mathbb{S}^3$ at different values of $\tilde{\eta} \in (-\tfrac{\pi}{2},\tfrac{\pi}{2})$. The two poles of this $\mathbb{S}^3$ sit on the vertical boundaries, and every point on it between those poles is itself an $\mathbb{S}^2$.}
	\label{Penrose_diagram_1}
\end{figure}

To understand the causal structure of de Sitter space, it is essential to bring both temporal and spatial infinities to a finite coordinate distance. 
We introduce a global conformal time $\tilde{\eta}$ as
\begin{equation}
d\tilde{\eta} = \frac{d \tau}{ \ell \cosh{\dfrac{\tau}{ \ell}}}~. 
\end{equation} With this change of these coordinates, metric~\eqref{global_dS_metric} becomes conformally flat:
\begin{equation}
ds^2 = \frac{ \ell^2}{\cos^2{\tilde{\eta}}}\Bigl( - \, d \tilde{\eta}^2 + d\Omega_3^2 \, \Bigr)~,
\end{equation} where $\tilde{\eta} \in  (-\tfrac{\pi}{2}, \tfrac{\pi}{2})$. The overall conformal factor diverges as $\tilde{\eta}$ approaches $ \pm \tfrac{\pi}{2}$, corresponding to the past and future temporal infinities of global de Sitter space, denoted by $\mathcal{I}^-$ and $\mathcal{I}^+$ respectively, where all null and timelike geodesics terminate. We present the Penrose diagram in Figure~\ref{Penrose_diagram_1}. Since conformal transformations preserve the  light cone structure, the causal structure of de Sitter is conformally equivalent to the Einstein static universe metric, except the range for $\tilde{\eta}$. 
From the causal structure of de Sitter space, we note a very important feature --- the presence of a cosmological event horizon. Any timelike curve can only access an incomplete portion of the full de Sitter spacetime, and there are always events that are forever out of causal contact. In fact, one can describe the region within the event horizon in an entirely static form.

To describe the static region of the de Sitter space, called the static patch, we consider the following parametrisation
\begin{align}\label{static_coord_param}
    x^0 &= \sqrt{ \ell^2-r^2\,} \, \sinh{\frac{t_s}{ \ell}}~, \nonumber \\
    x^1 &= r \, \sin{\theta_1}\cos{\theta_2}~, \nonumber \\
    x^2 &= r \, \sin{\theta_1}\sin{\theta_2}~, \\ \nonumber 
    x^3 &= r \, \cos{\theta_1}~, \\ \nonumber 
    x^4 &= \sqrt{ \ell^2-r^2\,} \, \cosh{\frac{t_s}{ \ell}}~,
\end{align} one obtains the following metric form
\begin{equation}\label{static_dS_metric}
ds^2 = - \biggl(1-\frac{r^2}{ \ell^2}\,\biggr) \, dt_s^2 + \frac{dr^2}{\biggl(1-\dfrac{r^2}{ \ell^2}\,\biggr)} + r^2 \, d\Omega_2^2~.
\end{equation} Here $t_s \in \mathbb{R}$ is time in the so-called static patch, $r \in (0, \ell)$, and $d\Omega_2^2$ is the metric of a two-dimensional unit sphere $\mathbb{S}^2$. Unlike the global coordinates~\eqref{global_coord_param} that cover the entire hyperboloid, the static patch coordinates, introduced by de Sitter in 1917~\cite{deSitter:1917}, cover the causal diamond formed by the intersection of the future lightcone of a past point and the past lightcone of a future point along an observer's worldline, since $x^0+x^4>0$ and $x^4>0$. 

\begin{figure}[t]
	\centering
	\includegraphics{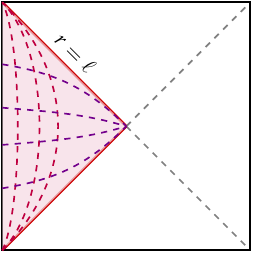}
\caption{The static patch corresponds to the shaded region. The dashed curved red lines represent slices of constant $r$, while the dashed curved purple lines correspond to slices of constant $t$.}
	\label{Penrose_diagram_2}
\end{figure}

In these coordinates, the surface found when $r$ approaches $\ell$, tends to the intersection of the future and the past de Sitter event horizon.  This is an infinite redshift surface. The accessible region for a single observer is bounded by the cosmological event horizon at $r= \ell$, see Figure~\ref{Penrose_diagram_2}. If $\Lambda$ is indeed constant in our own Universe, the size of our de Sitter horizon is approximately twenty billion light years in size. Within the static patch, the Killing vector $\partial_t$ is timelike away from the horizon and null on the horizon.
%Upon crossing horizon it becomes spacelike. 

\newpage
It is interesting to contrast the presence of a cosmological event horizon to the cosmological principle of homogeneity and isotropy introduced by Friedmann. The latter can be viewed as a way to dispense the theory from a distinguished observer. On the other hand, if observers are surrounded by an event horizon, perhaps their theory of physics should describe phenomena within their accessible region. 

The presence of a cosmological event horizon is not unique to de Sitter space. It is an instructive exercise to show that a spacetime of the type~\eqref{FLRW_metric}  with a monotonically increasing $a(t)$, and satisfying 
\begin{equation}
    \int\limits_{t_0}^\infty \frac{dt}{a(t)} < \infty~,
\end{equation} 
will expand at a sufficiently rapid rate that no observer will be able to causally access the entire spacetime.  Cosmological horizons do not just emerge in exponentially accelerating de Sitter universes, but in any model where the expansion is sufficiently rapid, i.e., whenever the scale factor grows faster than the Milne universe case~\eqref{Milne_metric}. What truly distinguishes de Sitter space is the existence of a static coordinate representation~\eqref{static_coord_param}, in which the metric becomes manifestly time-independent and the horizon appears as a coordinate singularity at a fixed radial distance. 
\newline\newline
{\textbf{De Sitter horizon thermodynamics.}} Although we will not focus on this in these lectures, we would like to mention an important conjecture about the nature of the de Sitter horizon. In \cite{Gibbons:1977mu}, drawing from ideas of black hole thermodynamics and the Bekenstein-Hawking proposal for the entropy of a black hole, Gibbons and Hawking postulated that the de Sitter horizon carries a quantum mechanical entropy, $\mathcal{S}$. This idea was further developed in \cite{fischler2000taking,Bousso:2000nf,Banks2003,Parikh:2004wh}. To leading order in the semiclassical regime, Gibbons and Hawking proposed that $\mathcal{S}$ is given by the area of the de Sitter horizon divided by $4G$, which can be expressed as follows
\begin{equation}
\mathcal{S} = \frac{3\pi c^3}{G \Lambda \hbar}~.
\end{equation}
We have reinstated the factor of $\hbar$ to emphasise this is a quantum mechanical entropy. For the currently measured value of $\Lambda$ in our own Universe, this takes the value $\mathcal{S}\approx 10^{122}$. This is the same puzzling number that often appears in the cosmological constant problem. Furthermore, the de Sitter horizon has a corresponding Hawking temperature that goes as $T_{\text{dS}} = (2\pi\ell)^{-1}$ \cite{Figari:1975km}. The analogous quantity for black holes, the Bekenstein-Hawking entropy $\mathcal{S}_{\text{BH}}$, is supported by significant evidence, starting with  the work of Strominger and Vafa \cite{Strominger:1996sh} who offered a microstate counting interpretation of $\mathcal{S}_{\text{BH}}$ for an exotic class of supersymmetric black holes. The AdS/CFT correspondence adds further evidence supporting the Bekenstein-Hawking relation --- large black holes in anti-de Sitter space are viewed as the bulk manifestation of highly entropic thermal states in the dual conformal field theory \cite{Witten:1998zw}. 

We do not have a complete microscopic understanding for $\mathcal{S}$, but there has been significant progress in recent years, including \cite{Anninos:2020hfj,Coleman:2021nor,Chandrasekaran:2022cip,Maldacena:2024spf,Collier:2025lux,Anninos:2026hia}. Having a more complete understanding of the underlying microscopic information content in a cosmological spacetime may offer a new perspective on when our current framework of effective field theory breaks down and how to complete it. As such, a microscopic understanding of $\mathcal{S}$ appears to be an important target for the theory quantum cosmology.

%\newpage 

%\begin{align} \frac{1}{r_1} &= \frac{\text{size of the largest telescope}}{\text{size of Universe}} \\ r_2 & = \frac{\text{size of ATLAS/LHC/...}}{\text{size of 10 TeV}}\label{eq:estimates}\end{align}

%More generally, the "asymptotical de Sitter spacetime" solution, $\Lambda>0$, near  $|\eta|\ll 1$ (Starobinsky, Fefferman-Graham)
%\begin{equation} \frac{ds^2}{l^2} = - \, \frac{d\eta^2}{\eta^2} + \frac{1}{\eta^2} \Bigl( \, g^{(0)}_{ij}(x) + \, \eta^2 g^{(2)}_{ij}(x) + \, \eta^3 g^{(3)}_{ij}(x) + \, ... \Bigr) dx^i dx^j \end{equation} small but finite convergence. 

\subsection{Gao-Wald Theorem}

In addition to the pure de Sitter spacetime it is important to understand the behaviour of classical solutions subject to perturbations. Moreover, the perturbation of a spacetime by positive-energy matter highlights some  differences between  cosmological event horizons and  black hole event horizons. In order to examine this, we first set up some important definitions.
A spacetime is said to satisfy the {null energy condition}
if, for all null $q^{\mu}$, one finds 
\begin{equation}
R_{\mu\nu}q^{\mu}q^{\nu}\geq0~,
\end{equation}
where $R_{\mu\nu}$ is the Ricci tensor of the spacetime. A spacetime
is said to satisfy the {null generic condition} if each null
geodesic contains a point at which 
\begin{equation}
k_{[\alpha}R_{\beta]\gamma\delta[\mu}k_{\nu]}k^{\gamma}k^{\delta}\neq0~,
\end{equation}
where $k^{\mu}$ denotes a tangent to the geodesic. Furthermore,
denote the {chronological future} of a point $p$ by $I^{+}\left(p\right)$,
which is the set of all points that can be reached from $p$ by a
timelike curve. The {chronological past}, $I^{-}\left(p\right)$,
similarly denotes the set of all points that can reach $p$ by timelike
curves. 

\begin{figure}[t]
	\centering
	\includegraphics{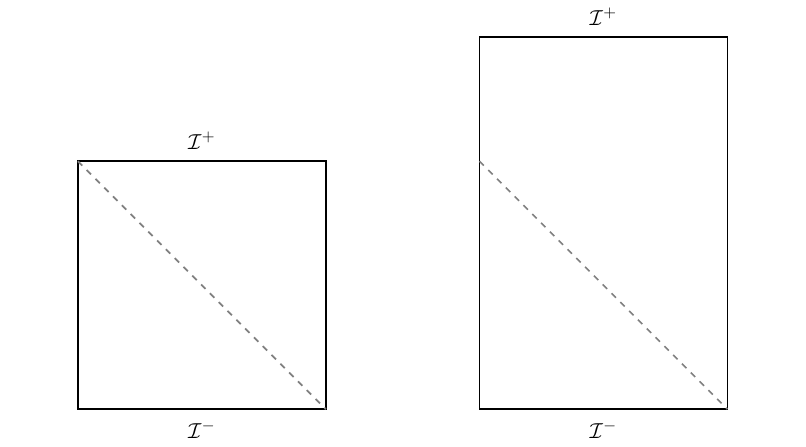}
\caption{{\it Left}: Penrose diagram of pure de Sitter. {\it Right}: “Stretched” Penrose diagram perturbed by positive-energy matter --.In response to positive-energy matter the de Sitter Penrose diagram stretches vertically. For pure de Sitter, the limiting null ray at the top left corner of $\mathcal{I}^{+}$ tracks back to the bottom right corner of $\mathcal{I}^{-}$. Since $\mathcal{I}^{-}$ is the initial Cauchy surface, it is not possible for an observer to have access to all timelike curves from any Cauchy surface in a finite amount of time, resulting from the failure of the spacetime to satisfy the null generic condition. Perturbing by positive-energy matter enhances the connectivity of the expanding spacetime, giving an observer access to an entire past Cauchy surface via timelike curves alone in a finite amount of time.}
	\label{fig:stretched}
\end{figure}

Consider a null geodesically complete spacetime satisfying the null
energy condition and the null generic condition. Suppose that in addition
this spacetime is globally hyperbolic (so that the initial value problem
is well-defined), with a compact Cauchy surface $\Sigma$. Gao and
Wald~\cite{Gao:2000ga} proved that for such a spacetime there exist Cauchy
surfaces, $\Sigma_{1}$ and $\Sigma_{2}$, with $\Sigma_{2}$ in the
chronological future of $\Sigma_{1}$, such that any point in the
chronological future of $\Sigma_{2}$ will have the entirety of $\Sigma_{1}$
in its chronological past. More formally, 
\[
\text{If }\Sigma_{2}\subset I^{+}\left(\Sigma_{1}\right),\text{then for all }p\in I^{+}\left(\Sigma_{2}\right)\text{ one finds }\Sigma_{1}\subset I^{-}\left(p\right).
\]
De Sitter spacetime satisfies all of the conditions of the proof, except
for the null generic condition. This is visualised in Figure~\ref{fig:stretched} ({\it left}): an observer arbitrarily far in the future will come arbitrarily
close to containing $\mathcal{I}^{-}$ in their chronological past.
Drawing a null ray from the top left corner of $\mathcal{I}^{+}$
hits the bottom right corner of $\mathcal{I}^{-}$ and so the entire
past light cone (including the null ray that defines this light cone)
contains $\mathcal{I}^{-}$. However, the chronological past, which
is made up of only timelike curves, does not. On the other hand, if
one were to perturb the spacetime with positive-energy matter such
that the null generic condition is satisfied, then all conditions
of the proof are met and a sufficiently late observer will be
able to view the entirety of $\mathcal{I}^{-}$, as shown in Figure~\ref{fig:stretched} ({\it right}). This has the interpretation that the Penrose diagram stretches
vertically in response to positive-energy matter. It encapsulates
the idea that matter generally contracts an expanding universe, bringing
more of the spacetime within the horizon of an observer. This is an
important concept for inflation, where one often describes the ``re-entering''
of superhorizon modes at the end of inflation.

The vertical stretching of the de Sitter Penrose diagram in response to positive-energy matter sharply contrasts the behaviour of a black
hole spacetime. While the effect of such matter on an expanding spacetime is to enhance the connectivity between the two sides of the cosmological
horizon, such that signals can be sent between previously causally-disconnected regions, for a black hole spacetime, such a matter perturbation further
de-correlates the two sides of the horizon. Effectively, positive-energy matter stretches the Penrose diagram of a black hole spacetime horizontally,
decreasing the connectivity. The formal statement stems from a theorem of Galloway \cite{galloway1995topology}, where it is shown for a spacetime obeying the null energy condition, that any two disconnected components of an asymptotically flat boundary are causally disconnected.  One should therefore be cautious when
pushing the analogy between cosmological and black hole horizons, as the dynamics of the two can differ significantly. 

\subsection{Asymptotically de Sitter geometries}

As a final note, we would like to mention the general late-time structure of solutions to Einstein's equations with $\Lambda>0$. In the works of \cite{Starobinsky:1982mr,fefferman1985conformal}, it is shown that the solutions admit a late-time, or equivalently small-$\eta$, structure given by
\begin{equation}\label{FG}
    ds^2= - \ell^2\frac{ d\eta^2}{\eta^2} +  \frac{ \ell^2}{\eta^2}  \bigl(  g_{ij}^{(0)} + \eta^2 g_{ij}^{(2)} + \eta^3 g_{ij}^{(3)}+ \ldots \bigr)dx^i dx^j~.
\end{equation}
Here the $g_{ij}^{(0)}$ is a Euclidean three metric, and $g_{ij}^{(2)} = R^{(0)}_{ij}-\tfrac{1}{4}R^{(0)} g_{ij}^{(0)}$ with $R_{ij}^{(0)}$ the Ricci tensor associated to $g_{ij}^{(0)}$. Finally, $g_{ij}^{(3)}$ is a symmetric tensor that is transverse and traceless with respect to $ g_{ij}^{(0)}$. All leading coefficients in the small-$\eta$ expansion of (\ref{FG}) are entirely fixed by the data $\left(g_{ij}^{(0)},g_{ij}^{(3)}\right)$. It is worth noting that there are residual diffeomorphisms that preserve the Fefferman-Graham gauge (\ref{FG}), under which the data $\left(g_{ij}^{(0)},g_{ij}^{(3)}\right)$ transform as conformal tensors. In \cite{Anninos:2010zf-ch04} an analysis of the asymptotic structure of diffeomorphism acting on  (\ref{FG}) is presented. A large portion of the solution space (\ref{FG}) will classically evolve to a big bang singularity in the far past. However, a small class of configurations, particularly ones for which the asymptotic data is close to that of global de Sitter space, will evolve to an $\mathcal{I}^-$ instead. This last statement is the classical stability theorem of four-dimensional de Sitter space proved by Friedrich in \cite{Friedrich:1986qfi}. A more general idea, known as the cosmological no-hair theorem, posits that a large class of initial data on some given Cauchy surface $\Sigma$, will classically evolve to (\ref{FG}) at sufficiently late times \cite{hawking1982supercooled,wald1983asymptotic,barrow1987cosmic,jensen1987inflation,East:2015ggf}. The precise relation between the asymptotic data $\left(g_{ij}^{(0)},g_{ij}^{(3)}\right)$ and the data along $\Sigma$ is not understood in any precise form.

\section{Quantum Field Theory in de Sitter space}

Having established the geometric structure of de Sitter spacetime, in this section we proceed to the study of quantum fields on a de Sitter background. We consider the problem through the lens of the maximal de Sitter isometries. In the semiclassical regime of interest, we treat de Sitter space as a fixed classical background and consider quantum matter fields propagating on it. This approximation amounts to neglecting gravitational backreaction, effectively working in the limit $G \to 0$, so that the geometry remains unperturbed by quantum fluctuations of the matter fields. The problem thus reduces to formulating quantum field theory on a rigid, curved, maximally symmetric spacetime.\footnote{For gravitational theories with de Sitter solutions populated with matter fields, the $G\to 0$ limit is delicate. Due to the compactness of the Cauchy surfaces, even when $G\to0$, there are still diffeomorphism constraints that need to be imposed \cite{Higuchi:1991tm} on  the space of states. Particularly, the $SO(1,4)$ de Sitter isometries are themselves coordinate transformations, so all states should be invariant under them. So more correctly, in this section we are really considering quantum fields on a rigid de Sitter background. Of course, there is still important information to be extracted from the rigid approximation, including many of the computations that are performed in inflationary cosmology. As such, though not formally precise when gravity is included in the picture, it is a well motivated problem.}

%\subsection{Unitary representation of dS$_d$}
A central conceptual issue in curved spacetime is the identification of particles and the organisation of the Hilbert space of states. In Minkowski space this structure is dictated by Poincar\'e invariance \cite{Wigner:1939cj}. In de Sitter space, the Hilbert space will be organised in terms of the $SO(1,d)$ isometry group. In particular, we assume that the Hilbert space of the quantum field theory, defined on a Cauchy slice (for instance $S^{d-1}$ in global coordinates or $\mathbb{R}^{d-1}$ in planar coordinates), is decomposed into unitary irreducible representations (UIRs) of $SO(1,d)$. In this framework, the notion of elementary field excitations is intrinsically tied to these UIRs, generalising the familiar Wigner classification of particle species in Minkowski spacetime.

In the following, we review the structure of de Sitter isometries and their unitary representations, which will be useful for constructing and interpreting quantum fields on the background. Given the non-compact, semisimple nature of the $SO(1,d)$ isometry group, unless trivial, the UIRs will all be infinite dimensional.
We will first study the two-dimensional de Sitter space, which has the isometry group $SO(1,2)$, due to its simplicity and its particular similarity to $SO(1,4)$. The problem of UIRs for semisimple non-compact Lie groups was solved by Harish-Chandra in generality, which $SO(1,d)$ is an example of. The reader is encouraged to go to the following references for the detailed derivation of UIRs of $SO(1,d)$ group \cite{Dobrev:1977qv,Sun_2024}, and for a physics-based analysis with a particular emphasis on $SO(1,2)$ see \cite{Anninos:2023lin}.

\subsection{$SO(1,2)$ UIRs}

Here we present the UIRs for the $SO(1,2)$ group, which at the level of the $\mathfrak{so}(1,2)$ Lie algebra is isomorphic to $\mathfrak{sl}(2,R)$, namely $2\times 2$ matrices with unit determinant and real valued entries.\footnote{When in doubt, the reader should assume we are discussing $SO^+(1,2)$, the component of $SO(1,2)$ continuously connected to the identity or, if specified, the double cover of $SO^+(1,2)$ which is $SL(2,\mathbb{R})$.} The group is generated by three ladder type generators $\{L_0,L_\pm\}$ with algebra
\begin{equation}
    [L_0,L_\pm]=\mp L_\pm~, \quad [L_+,L_-]=2L_0~. 
\end{equation}
The ladder operators obey the following Hermiticity conditions
\begin{equation}
    L_\pm^\dagger= L_\mp~, \qquad L_0^\dagger=L_0~.
\end{equation}
We have also have a Hermitean quadratic Casimir operator, $C$,  which commutes with all the generators of $SO(1,2)$. Its explicit form is given by
\begin{equation}
    C=\frac{1}{2}(L_+L_-+L_-L_+)-L_0^2 \equiv\lambda(1-\lambda)~, \quad \lambda\in\mathbb{C}~.
\end{equation}
Unitarity enforces the reality of $C$, which in turn implies that $\lambda$ is either real valued or of the specific form $\lambda=\tfrac{1}{2}+i\nu$, with $\nu\in\mathbb{R}$. We now discuss the three branches of $SO(1,2)$ UIRs, which further restrict the allowed values of $\lambda$.
\newline\newline
{\textbf{Discrete series.}} The discrete series irreps of de Sitter ($\mathcal{D}_\lambda^\pm$) can be defined by taking an eigenstate of $L_0$, namely $L_0\ket{\lambda}=\mp\lambda\ket{\lambda}$, with $\lambda\in\mathbb{Z}^+$. We have two options here, either to consider this state to be annihilated by $L_+$ or $L_-$, and these correspond to the $\pm$ branches of the discrete series $\mathcal{D}_\lambda^\pm$. Let us consider $\mathcal{D}_\lambda^-$, such that we take $\ket{\lambda}$ to obey $L_-\ket{\lambda}=0$. Acting on this state with powers of $L_+$ gives other states within the UIR, namely $L_+^n\ket{\lambda}=\ket{n,\lambda}$, with $n\in\mathbb{N}$. We note that $L_0\ket{n,\lambda}= -(\lambda+n)\ket{n,\lambda}$. Similarly we can define the other branch of the discrete series using an eigenstate of $L_0$ which is annihilated by $L_+$. This gives us a bounded, infinite dimensional unitary irreducible representation. This has an analogue in four-dimensional de Sitter space,  where one has the additional label  corresponding to the intrinsic spin of the state coming from the $SO(3)$ subgroup of $SO(1,4)$. For $SO(1,2)$, the spin would be associated to  an $SO(1)$ subgroup of $SO(1,2)$ which is trivial. For the double cover of $SO(1,2)$, i.e. $SL(2,\mathbb{R})$, the subgroup becomes $O(1) \cong\mathbb{Z}_2$, which can be used to produce a further  $\mathbb{Z}_2$ grading of UIRs, and one can also have half-integer values for the $L_0$ eigenvalues. The latter are the analogues of fermionic representations for the double cover of $SO(1,2)$.
\newline\newline
{\textbf{Continuous series.}} In addition to discrete series we have other irreps which can be studied by letting $\lambda$ to take non-integer values. Consider a state $\ket{\lambda,n}$ such that $L_0\ket{n,\lambda}=n\ket{n,\lambda}$, where $n\in \mathbb{Z}$.  Computing the norm of the state $L_+\ket{n,\lambda}$, one finds
\begin{equation}
    \bra{n,\lambda}L_+^\dagger L_+\ket{n,\lambda}>0 \quad \forall\quad n\in\mathbb{Z}~ \quad\quad
    \implies \quad\quad  \bra{n,\lambda}L_-L_+\ket{n,\lambda}>0~,
\end{equation}
where we have used $L_+^\dagger=L_-$. Similarly, 
\begin{equation}
    \bra{n,\lambda}L_+L_-\ket{n,\lambda}>0 \quad \forall\quad n\in\mathbb{Z}~.
\end{equation}
We can combine the above two conditions to impose constraints on $\lambda$,
\begin{equation}
    \bra{n,\lambda}L_-L_++L_+L_-\ket{n,\lambda}= 2\bra{n,\lambda}(C+L_0^2)\ket{n,\lambda}>0~  \quad \implies \quad \lambda(1-\lambda)>0~.
   % &\implies n^2+\lambda(1-\lambda)>0 \quad\forall\quad n\in\mathbb{Z}~,\\ \n
\end{equation}
Taking $\lambda\in\mathbb{C}$, we get the two branches 
\begin{align}\nonumber
    \lambda&\in(0,1)~,  \qquad  \hspace{1cm}&&: \,\, \text{Complementary series}\\\nn 
    \lambda&=\frac{1}{2}+i\nu~,\quad \nu\in\mathbb{R} &&:\,\,\text{Principal series}
\end{align}
The principal, complementary, and two discrete series, along with the trivial representation, constitute  the complete set of UIRs for $SO(1,2)$. The single particle Hilbert space of a quantum field theory on a rigid de Sitter background can be partitioned in a direct sum of these three types of UIRs, or their corresponding double-cover for the case of fermions. 

%These UIRs themselves inform us about the single particle states in de Sitter space. Again, for the double cover $n$ can take half-integer values.

\subsection{$SO(1,4)$ UIRs}

The four-dimensional de Sitter space has analogous UIRs, which we describe below, along with the additional $SO(3)$ spin label, $s\in\mathbb{N}$, which corresponds to the intrinsic spin of the particle. The eigenvalue of quadratic Casimir is modified because of the spin
\begin{align}
    C_2=\Delta(3-\Delta)-s(s+1)~.
\end{align}
We have traded $\lambda$ for $\Delta$ to distinguish the two-dimensional and four-dimensional cases. In four-dimensions, we have an additional quartic Casimir operator, $C_4$, justifying the subscript in $C$. Although we do not use it, we mention its eigenvalue for the sake of completeness
\begin{align}
    C_4=s(s+1)(\Delta(3-\Delta)-2)~.
\end{align}
The three series of UIRs for four-dimensional de Sitter space can be summarised as follows:
\begin{itemize}
    \item[] \textbf{Principal series ($\pi_s^\Delta$)}: These mimic the 2-dimensional case. However,  the real part of $\Delta$ is different, namely $\Delta=\frac{3}{2}+i\nu$ with $\nu\in\mathbb{R}$. The real part depends on the dimension of the de Sitter spacetime and for general $d$ is given $\Delta=\frac{d-1}{2}+i\nu$. Physically, this representation is associated to massive (in units of $\ell$) particles in de Sitter space. 
    \item[] \textbf{Complementary series ($\tilde{\pi}_s^\Delta$)}: In four-dimensions, the complementary series takes the range $\Delta\in(1,2)$ for higher spins $s\in\mathbb{Z}^+$. The $s=0$ complementary series acts differently, admitting the larger range $\Delta\in(0,3)$. Physically, this representation is associated to light particles in de Sitter space.
    \item[] \textbf{Discrete series/Exceptional series ($\mathcal{D}^\pm_{s,t}$)}: Here as well, scalar UIRs behave differently than higher spin UIRs. For $s=0$, $\Delta=3+k$, where $k\in\mathbb{N}$. These are known as the type I exceptional series of $SO(1,4)$.
    %the states created using these primary states are reducible and should be projected out and therefore are not discrete series in true sense.\\
     For $s\in\mathbb{Z}^+$, we have a parameter $t\in\{0,1,2,..,s-1\}$ called the depth which fixes the value of $\Delta=2+t$. Physically, the representation with $t=s-1$ is associated to strictly massless gauge fields, while the other depths are known as partially massless gauge fields. The latter have no Minkowski space analogue.
\end{itemize}
We can also study the fermions in de Sitter space by considering the double cover $SO(1,4)$. The UIRs for fermions also have principal and discrete series, however, the complementary series is absent for fermions \cite{thomas_1940,ottoson1968classification,schwarz1971unitary}. For a physics perspective of the fermionic representations at higher spin see \cite{Letsios:2023qzq,Anninos:2025mje}. 
\newline\newline
{\textbf{Tensor products of $SO(1,2)$, briefly.}} A more complete understanding, coding multiparticle states, necessitates an understanding of the tensor products of these UIRs also. These UIRs are realised by the single particle Hilbert space, and we further need to study the tensor products of the UIRs to describe the multiparticle Hilbert space. We won't go into the details of the structure but just state, schematically, how tensor products of $SO(1,2)$ UIRs of different series \cite{repka1978tensor,repka1979tensor} can be decomposed in terms of the direct sum of UIRs (see \cite{Penedones:2023uqc} for a physics based discussion).
\begin{itemize}
    \item[$\bullet$] $\mathcal{D}_{\lambda_1}^\pm \otimes \mathcal{D}_{\lambda_2}^\pm = \sum_\lambda\oplus \mathcal{D}_\lambda^\pm$.  This means that if we are in any one branch of discrete series, then we do not find other UIRs in the tensor product. 
    \item[$\bullet$] $\mathcal{D}_{\lambda_1}^+ \otimes \mathcal{D}_{\lambda_2}^- = \int_\nu\pi_\nu\oplus\sum_\lambda \oplus\mathcal{D}_\lambda^\pm$. Physically, this means that even if we start with gauge fields whose elementary excitations are described by $\mathcal{D}_\lambda^\pm$, their multiparticle Hilbert space can encode the principal series that ordinarily describe the single particle Hilbert space of massive fields.
    \item[$\bullet$] $\pi_{\nu_1}\otimes\pi_{\nu_2}=\int_\nu\pi_\nu\oplus \sum_{\lambda_1}\oplus \mathcal{D}_{\lambda_1}^+\oplus \sum_{\lambda_2}\oplus \mathcal{D}_{\lambda_2}^-$. This is the converse of the previous observation. Physically, it means that the multiparticle Hilbert space of heavy particles contains states that we would associate to elementary excitations of massless gauge fields.
\end{itemize}
Generally speaking, complementary series do not appear in tensor product decomposition of non-complementary UIRs, in this sense they are somewhat isolated from the other series.

\subsection{Harish--Chandra Characters}

A useful way to characterise the information contained in a unitary irreducible representation (UIR) of the de Sitter group is through its {Harish--Chandra character} \cite{HarishChandra1955OnTC}. Let $\pi_\lambda$ be a representation of the group $G$ on a Hilbert space $\mathcal H_\lambda$. For two-dimensional de Sitter space, $G = SO(1,2)$. The character is formally defined as the trace
\begin{align}
\chi_\lambda(t) = \mathrm{Tr}_{\mathcal{H}_\lambda} \, e^{i t D}~, \quad\quad t\in\mathbb{R}~,
\end{align}
where $D$ denotes the Hermitean generator of the $\mathfrak{so}(1,1)$ one-dimensional non-compact Cartan subalgebra of $\mathfrak{so}(1,2)$. Equivalently, $e^{i tD}$ generates the $SO(1,1)$ (hyperbolic) subgroup of $SO(1,2)$. Since the group is non-compact, this trace is not absolutely convergent and must be understood in a generalised sense. The character is therefore defined as a {conjugation-invariant distribution} on the group. It depends only on the conjugacy class of the element $e^{itD}$. %Two irreducible representations are equivalent if and only if they have the same character. 
\newline\newline
{\textbf{Continuous series characters.}} For the principal series, or complementary series UIRs of $SO(1,2)$, labeled by $\lambda$, the character evaluated on hyperbolic elements $e^{i tD}$ takes the form
\begin{align}
\chi_\lambda(t)
=
\frac{e^{-t\lambda} + e^{-t(1-\lambda)}}{1 - e^{-t}}~.
\end{align}
The denominator arises universally from the group structure and reflects the fact that we are evaluating the trace on hyperbolic conjugacy classes, while the numerator encodes the representation parameter. 
\newline\newline
{\textbf{Discrete series characters.}} For the discrete series URIs, $\mathcal{D}_\lambda^\pm$, the character evaluated on hyperbolic elements $e^{i tD}$ takes the form
\begin{align}
\chi_{\lambda}(t) = \frac{e^{-t\lambda}}{1 - e^{-t}}~.
\end{align}
We note that the discrete series character is structurally different from the one for the continuous series.

\subsection{Scalar quantum field on rigid de Sitter space}

At this stage we would like to associate some of the UIRs of $SO(1,d)$ with free quantum fields in $d$-dimensional de Sitter space. To illustrate this we will  consider a minimally coupled massive scalar quantum field, $\phi$, on a four-dimensional rigid de Sitter background. The action is given by
\begin{align}
    S= -\frac{1}{2}\int d^4x\sqrt{-g}(g^{\mu\nu}\partial_\mu\phi\partial_\nu\phi +m^2\phi^2)~.
\end{align}
The Klein-Gordon equation of motion for the scalar field is given by
\begin{align}
    -\frac{1}{\sqrt{-g}}\partial_\mu(\sqrt{-g}g^{\mu\nu}\partial_\nu\phi)=m^2\phi^2~.
\end{align}
The operator on the left hand side in the above equation is nothing but the Casimir operator, represented in terms of differential operators. The de Sitter generators in this representation are the Killing vector fields under the Lie bracket. Thus, we have
\begin{equation}
    {C_2\phi}=-\frac{l^2}{\sqrt{-g}}\partial_\mu(\sqrt{-g}g^{\mu\nu}\partial_\nu\phi)= m^2 \ell^2\phi~ \quad\quad 
    \implies  \quad\quad {\Delta(3-\Delta)}= m^2 \ell^2~.
\end{equation}
This allows us to relate $\Delta$ to the mass, $m$, of the scalar field 
\begin{align}\label{scalar mass}
    \Delta=\frac{3}{2}\pm\sqrt{\frac{9}{4}-m^2\ell^2}~.
\end{align}
% For the case of free massive  higher spins fields, this gets modified to
% \begin{align}\label{conf_high_spin}
%     \Delta=\frac{3}{2}\pm\sqrt{\frac{9}{4}-(m^2l^2+(s-2)(s+1))}~.
% \end{align}
%Note that we don't get (\ref{scalar mass}) by simply putting $s=0$ in the above expression and one needs to be careful about it.
We can associate different series representations that we found before to different range of particle masses using the above relations. In particular:
\begin{itemize}
    \item[] \textbf{Heavy de Sitter particles:}  If $m^2\ell^2\ge\tfrac{9}{4}$, $\Delta$ becomes complex and the imaginary part is related to the mass $m$. From this it will follow, upon quantisation, that single particle states of a heavy  (in units of $\ell$) scalar field in de Sitter space furnish the principal series of $SO(1,4)$.
    \item[] \textbf{Light de Sitter particles:} Light particles in de Sitter space, satisfying $m^2\ell^2 \in (0,\tfrac{9}{4})$, have $\Delta\in (0,3)$. These furnish the complementary series of $SO(1,4)$.
    \item[] \textbf{`Massless' de Sitter particles:} The strictly massless scalar field yields $\Delta=3$. This corresponds to the boundary value of the exceptional series. It is subtle to quantise a massless minimally coupled free scalar in de Sitter space in a de Sitter invariant way. One possibility is to gauge the shift symmetry and try to quantise the ensuing theory \cite{Folacci:1992xc,Anninos:2023lin}. Alternatively, adding small interactions can cure some of the infrared pathologies of the free massless field \cite{Starobinsky:1994bd-ch04,Gorbenko:2019rza-ch04}.
    %It is unclear what field theory corresponds to the other representations in the exceptional series.
\end{itemize}
For spin-$s$ free fields on a rigid four-dimensional de Sitter background, the above range gets modified in the following way (see for example \cite{Baumann:2017jvh}):
\begin{itemize}
    \item[] \textbf{Heavy spin-$s$ de Sitter particles:} The spinning principal series UIR satisfy $\Delta(3-\Delta)\geq\frac{9}{4}$ for heavy fields with a mass term given by
    \begin{equation}
        m^2\ell^2\geq (s-\frac{1}{2})^2~, \quad s\in\mathbb{Z}^+~.
    \end{equation}
    \item[] \textbf{Light spin-$s$ de Sitter particles:} The spinning complementary series UIR satisfy $2<\Delta(3-\Delta)<\frac{9}{4}$ with real $\Delta$. They are furnished by light spin-$s$ particles whose mass is in the range 
    \begin{equation}
        s(s-1)<m^2\ell^2< (s-\frac{1}{2})^2~, \quad s\in\mathbb{Z}^+~.
    \end{equation}
    We note that the mass parameter $m$ obeys a lower bound that is non-vanishing. This is known as Higuchi bound \cite{Higuchi:1986py}. Below this value of $m$ there are only a discrete set of isolated values that are unitarily allowed. There is no Minkowskian counterpart to this phenomenon. 
    \item[] \textbf{de Sitter gauge fields:} The spin-$s$ discrete series UIR satisfy $\Delta(3-\Delta)=(1-t)(2+t)$, where the depth parameter satisfies $t=0,\ldots,s-1$. The mass parameter reads
    \begin{equation}
        m^2\ell^2=(s-1-t)(s+t)~, \quad s\in\mathbb{Z}^+~.
    \end{equation}
At depth $t = s-1$, the fields are totally massless, carrying two locally propagating degrees of freedom. These fields resemble massless higher-spin gauge fields in Minkowski spacetime. At the free level, they are described by the Fronsdal action \cite{Fronsdal:1978rb}. For the other depths, $0 \le t < s-1$, the fields have $n_{s,t} = 2(s-t)$ locally propagating degrees of freedom. This contrasts with massive fields in Minkowski spacetime, massive fields of spin-$s$ have $2s+1$ degrees of freedom.  These intermediate cases, for $t\neq s-1$ are referred to as {partially massless fields}. The totally massless spin-1 case was analysed in \cite{RiosFukelman:2023mgq}, and the totally massless spin-2 case was analysed in \cite{Higuchi:1991tn}.
\end{itemize}
%more than 2 but fewer than $2s+1$

%\noindent Note that in terms of $\Delta(3-\Delta)$ the separation between the different series representation for higher spins ($s=1,2,..$) is independent of the particular spin-$s$ value. However, since the relation between $\Delta$ and $m$ depends on the value of $s$, the series separation in terms of $m$ becomes $s$ dependent.\\

\subsection{Canonical Quantisation}

At this stage we have related the labels of the UIR, ($\Delta,s$), to the more standard free field parameters $(m,s)$. We now consider the canonical quantisation of the quantum field theory. To do so, we promote $\phi$ to a field operator, and similarly for its momentum conjugate $\pi=\tfrac{\partial\mathcal{L}}{\partial\dot\phi}$. As usual, one imposes the equal time commutation relation. In global coordinates
\begin{equation}\label{comm}
    [\hat{\phi}(t,\Omega),\hat{\pi}(t,\Omega')]=i\delta(\Omega-\Omega')~,
\end{equation}
where $\Omega$ denotes the coordinates on the round unit three-sphere. The field operator $\hat{\phi}(t,\Omega)$ can be expanded in terms of harmonic functions on 3-sphere 
\begin{equation}
    \hat{\phi}(t,\Omega)={\sum}_{\vec{L}} \,f_{\vec{L}}(t)  Y_{\vec{L}}(\Omega) \hat{a}_{\vec{L}}+\text{h.c.} ~,
\end{equation}
where $Y_{\vec{L}}(\Omega)$ are the orthonormal spherical harmonics on the three-sphere. 

The mode functions, $f_{\vec{L}}(t)$, satisfy the Klein-Gordon equation, with a normalisation fixed upon imposing (\ref{comm}). In order to fully fix $f_{\vec{L}}(t)$, we further require that they obey a type of vacuum condition that has been argued to be natural in the context of de Sitter space from a variety of perspectives \cite{Chernikov:1968zm,Bunch:1978yq-ch04,hartle1983wave,Schomblond:1976xc,Mottola:1984ar-ch04,Allen:1985ux-ch04}. In global coordinates, the choice is such that upon analytically continuing $t$ to the lower Euclidean hemisphere, the functions $f_{\vec{L}}(t)$ remains smooth. This gives rise to a de Sitter invariant vacuum state $|0\rangle$, that is annihilated by the full set of $\hat{a}_{\vec{L}}$ operators. The single particle Hilbert space is created by acting once on $|0\rangle$ with the creation operators, $\mathcal{H}_\text{single particle} = \{a^\dagger_{\vec{L}}\ket{0},\, \forall \, \,\vec{L}\}$.

%We define the vacuum state to be the one annihilated by the creation operator $\hat{a}_{\vec{L}}\ket{0}=0$. 
\subsection{Late-time structure}

As a final comment, it is instructive to study the late-time structure of quantum field theory on a rigid de Sitter spacetime. One motivation for this stems from the interpretation of the cosmic microwave background radiation in terms of late-time quantum fluctuations of the inflaton field during the primordial inflationary era. Since we are concerned with the local structure of a constant and late-time slice of de Sitter space, we can work in the planar coordinates (\ref{flat_dS_metric_conformal_time}) of de Sitter space for simplicity. The Killing vectors of de Sitter space in these coordinates take the following form 
\begin{align}\nonumber
    \text{Spatial translations:}& \qquad P_i= \partial_i~, \\\nn
    \text{Dilatation:}& \qquad D=\eta\partial_\eta+x^i\partial_i~,\\\nn
    \text{Spatial Rotations:}& \qquad J_{ij}=x_i\partial_j-x_j\partial_i~,\\\nn
    \text{Special conformal transformations:}&\qquad K_i= 2x_iD+(\eta^2-\delta_{jk}x^j x^k)P_i~.  
\end{align}
Upon projecting these onto $\mathcal{I}^+$, these act as conformal transformations of the Euclidean metric on $\mathbb{R}^3$ whose coordinates we denote by $x^i$. This is a manifestation of the fact that the $SO(1,4)$ is the Euclidean conformal group in three-dimensions. The structure implies that the free scalar field operator admits the following late-time expansion
\begin{align}
   \hat{\phi}(\eta,{x}^i) = \hat{\alpha}_\Delta({x}^i)\eta^\Delta\left( 1+ \ldots \right)+\hat{\beta}_{{\Delta}}({x}^i)\eta^{3-\Delta}\left(1+\ldots \right)~,
\end{align}
which can be confirmed directly by solving the Klein-Gordon equation. The late-time operators, $\hat{\alpha}_\Delta({x}^i)$ and $\hat{\beta}_{{\Delta}}({x}^i)$, transform as conformal primary operators of $SO(1,4)$. 

If $\Delta$ lies in the principal series, such that $\Delta=\tfrac{3}{2}+i\nu$ with $\nu\in\mathbb{R}$, then $\bar{\Delta}\equiv3-\Delta = \Delta^*$ and the Hermiticity condition on $\hat{\phi} = \hat{\phi}^\dag$ implies $\hat{\beta}=\hat{\alpha}^{\dagger}$. Moreover,  they must satisfy the commutation relation 
\begin{equation}\label{princ_conf_operator}
    [\hat{\alpha}_\Delta({x}^i),\hat{\alpha}_\Delta^\dagger({y}^i)]=\delta^3({x}^i-{y}^i)~.
\end{equation}
Thus, the late-time operators $\hat{\beta}_\Delta$ and $\hat{\alpha}_\Delta$ take the form of creation and annihilation operators at the boundary. Nonetheless,  they do not annihilate $|0\rangle$. If $\Delta$ lies in the complementary series, then $\Delta$ is real and it follows that $\hat{\alpha}_\Delta(x^i)$ and $\hat{\beta}_\Delta(x^i)$ are independent Hermitian operators. They now obey $[\hat{\alpha}_\Delta({x}^i),\hat{\beta}_\Delta({y}^i)]=i\delta^3({x}^i-{y}^i)$. 
\newline\newline
{\textbf{{Late-time correlation functions, briefly.}}} We would like to use these conformal operators to study the correlation function at the late-time boundary. As our quantum state, we select the de Sitter invariant vacuum state $\ket{0}$. The two point function of the conformal operators, $\hat{\alpha}_\Delta({x}^i)$ and $\hat{\beta}_{{\Delta}}({x}^i)$,  is given by the standard two-point correlation function of a three-dimensional Euclidean conformal field theory. For instance,
\begin{equation}\label{Galpha}
    \bra{0}\hat{\alpha}_\Delta({x}^i)\hat{\alpha}_\Delta({y}^i)\ket{0}= c_\Delta \left(\frac{1}{|{x}^i-{y}^i|^2} \right)^{\Delta}~,
\end{equation} %G_\Delta^\text{free}({x}^i,{y}^i)=
where $c_\Delta \in \mathbb{C}$ is a normalisation constant. One interesting feature about the above relation is that if we have light scalar fields, such that $\Delta$ lies in the complementary series, the correlation function can decay very slowly as a function of the spatial separation. This is because $\Delta$ can approach zero in this case. Therefore, the correlation functions at late-times need not decay even though the spacetime is expanding rapidly, and the fluctuations can be imprinted near the late-time boundary. This mechanism is essential for inflation. The slow spatial decay of correlation functions for  light fields implies that  cluster decomposition is not generally obeyed in de Sitter space. For heavy fields, $\Delta=\tfrac{3}{2}+i\nu$, and one has an oscillatory correlation function with polynomially decaying magnitude.

Another interesting property of these late-time operators is that when acting on any normalised state, $\ket{\Psi}$, belonging to the quantum field theoretic Hilbert space, $\mathcal{H}_{\text{QFT}}$, they produce a non-normalizable state that is no longer in $\mathcal{H}_{\text{QFT}}$. This can be easily seen by the action of operator $\hat{\alpha}_\Delta(x^i)$ on the de Sitter invariant vacuum state. Let us take $\Delta$ to be in the complementary series for simplicity, then by
% \begin{align}
%     \bra{0}\hat{\alpha}_\Delta({x}^i)\hat{\alpha}_\Delta({y}^i)\ket{0} = c_\Delta  \left(\frac{1}{({x}^i-{y}^i)({x}^i-{y}^i)} \right)^{\Delta}~.
% \end{align}
% %where we have used the commutation relation (\ref{princ_conf_operator}). 
taking ${x}^i\rightarrow{y}^i$ in (\ref{Galpha}), we see that the norm of the state $\hat{\alpha}_\Delta({y}^i)\ket{0}$ diverges.
%which cannot be cured using counterterms in the Lagrangian.
\newline\newline
{\textbf{dS/CFT?}} Although there is a natural $SO(1,d)$ Euclidean conformal structure residing at $\mathcal{I}^+$,  there are important differences to the usual Euclidean conformal field theory analysis, as well as the $SO(1,d)$ structure near the boundary of Euclidean anti-de Sitter space.\footnote{Somewhat amusingly, at the level of Lie algebras, the symmetries of two-dimensional Lorentzian and Euclidean anti-de Sitter space are the same as those of two-dimensional de Sitter space. This was explored in \cite{Anninos:2019oka}.} For instance, in anti-de Sitter space, we have one boundary conformal operator corresponding to a bulk field operator, while in de Sitter we have two. Also, in Euclidean signature operators do not ordinarily have non-trivial commutation relations such as the ones in (\ref{princ_conf_operator}) satisfied by $\hat{\alpha}_\Delta({x}^i)$ and $\hat{\beta}_\Delta({y}^i)$. Moreover, the range of allowed $\Delta$ that appear in de Sitter space, is different to the one we ordinarily study in unitary conformal field theory. The reason is that we are imposing different notions of unitarity in each case. Ultimately, all this is linked to the fact that although the de Sitter isometries are isomorphic to the symmetries of a lower dimensional Euclidean conformal field theory, the underlying de Sitter theory is Lorentzian, coding non-trivial time-evolution and causal structure. Whatever the relation of physical quantities in $d$-dimensional de Sitter space are to $(d-1)$-dimensional conformal field theory \cite{Strominger:2001pn}, it must encode all these structures. Adding gravity to the mix only makes things more intricate ---  in gravity we have to view the de Sitter isometries as gauge redundancies \cite{Higuchi:1991tm}. Moreover, correlation functions labelled by specific spacetime points are not coordinate invariant, or even locally field redefinition invariant for that matter.

\section{Canonical quantum gravity}

In this section, we consider the canonical picture of quantum gravity pioneered by DeWitt in \cite{dewitt1967quantum,DeWitt:1967ub}. General relativity is usually presented in a way that unifies space and time. 
Nevertheless, it is often valuable to formulate a physical problem in terms of the initial value problem where the initial configuration of fields together with their associated momenta are specified. 
We discuss phase space approach of Arnowitt, Deser, and Misner (ADM) \cite{arnowitt1960canonical} to general relativity, and develop the quantum Wheeler-DeWitt equation.

\subsection{ADM formulation}

The canonical approach of ADM applies to a manifold, $\mathcal{M}$, that admits a foliation $\mathbb{R} \times \Sigma   \rightarrow \mathcal{M}$, where $\Sigma$ is a three-dimensional spacelike manifold, and $\mathbb{R}$ represents the time direction. For the sake of simplicity, we will assume that $\Sigma$ is closed throughout this section. More specifically, we will consider a $(3+1)$-decomposition of a four-dimensional pseudo-Riemannian spacetime $\mathcal{M}$ with coordinates $(t,x^i)$. The Greek indices take values $\mu\in\{0,1,2,3\}$, while the Latin indices take values $i\in\{1,2,3\}$. The successive hypersurfaces on which we will describe the geometry are  described by successive values of  the parameter $t$. However, a theory with a preferred foliation will break the manifest diffeomorphism invariance of the underlying theory, and we must recover this later.

Let $g_{\mu\nu}$ denote the spacetime metric on the four-dimensional manifold $\mathcal{M}$. Then, assuming the above foliation we choose a unit normal $n^\mu$ to the spacelike surface $\Sigma$. We may project the metric onto components parallel and perpendicular to the surface
\begin{equation}
ds^2 = -N^2 dt^2+h_{ij}(dx^i+N^i dt)(dx^j+N^j dt)~,
\end{equation}
here $h_{ij}(t,x^i)$ is the induced metric on $\Sigma$. $N(t,x^i) \in \mathbb{R}^+$ is called the lapse function and $N^i(t,x^i) \in \mathbb{R}^3$ is known as the shift vector. We note that $N(t,x^i)$, and $N^i(t,x^i)$ are not physical fields in the same sense as the spatial metric $h_{ij}$, instead they encode gauge freedom of how the spacetime is built by stacking spatial slices. The lapse and shift fields tell us how one moves from one $\Sigma$ at $t$ to another $\Sigma$ at $t+dt$.
Concretely, the lapse $N(t,x^i)$ captures the lapse of proper time along the normal direction between hypersurfaces at $t$ and $t+dt$. The shift, $N^i(t,x^i)$, captures how spatial coordinates slide tangentially along the slice as you go from the hypersurfaces at $t$ and $t+dt$. The unit timelike vector, $n^\mu$, normal to the hypersurfaces $\Sigma$ is determined by the lapse and shift function as follows
\begin{equation}
n^\mu = {N}^{-1}\left(1, - {N^i}\right)~.
\end{equation}
One is also lead to define a time flow vector
\begin{equation}
t^\mu = N n^\mu + N^\mu~,
\end{equation}
where $N^\mu = (0,N^i)$. Again we will only explicitly deal with spacelike embeddings such that $t^\mu$ is timelike. From this we can deduce that $N$ is nowhere vanishing and since we will choose to deal with future directed foliations we will also require $N$ to be everywhere positive.

Another quantity of importance is the extrinsic curvature, $K_{ij}$ of the three-geometries. Expressed in terms of the lapse and shift functions, and the induced metric $h_{ij}$ on $\Sigma$ one has
\begin{equation}\label{extrinsic curvature}
K_{ij} \equiv \frac{1}{2N} \left( \nabla_j N_i +\nabla_i N_j-\frac{\partial h_{ij}}{\partial t}\right)~,
\end{equation}
where $\nabla_i$ is the covariant derivative intrinsic to the 3-geometry.
%\textcolor{ForestGreen}{We will treat the 3-geometries of these hypersurfaces and the 4-geometry that fills in between these lamination. Then you should not how this breaks explicitly diffeomorphism invariance and we select out a preferred coordinate called $t$.}

Now we may begin to reformulate the Einstein-Hilbert action in terms of the foliation variables. Our original Einstein-Hilbert action
\begin{equation}\label{action EH}
S_{\text{EH}} = \frac{1}{16 \pi G_N}\int d^4x \sqrt{-g}(R-2 \Lambda)~,
\end{equation}
can be rewritten in terms of the foliation variables as follows
\begin{equation}\label{EH 3+1 variables}
S_{\text{EH}} = \frac{1}{16 \pi G_N} \int d^4x\, N \sqrt{h}\,\left(R^{(3)}-2\Lambda+K_{ij}K^{ij}-K^2\right)~,
\end{equation}
%\begin{equation}\label{EH 3+1 variables}
%\frac{1}{16 \pi G} \int d^4x\, N \sqrt{h}\,[R-2\Lambda+(\text{Tr }\bold{K}^2)-(\text{Tr } \bold{K})^2]~,
%\end{equation}
where $R^{(3)}$ represents the three-dimensional scalar curvature invariant associated to $h_{ij}$, and $K = h^{ij} K_{ij}$. The action $S_{\text{EH}}$ is independent of $\dot{N}$ and $\dot{N}^i$. As such, the lapse and shift have no conjugate momenta. Their role is to implement the constraints associated to the invariance under coordinate transformations. 

Next we define the phase space variables of our theory. The conjugate momentum to the dynamical metric $h_{ij}$ is obtained by varying the action with respect to $\dot{h}_{ij}$. One finds
\begin{equation}\label{Conjugate momentum}
\pi^{ij} = \frac{\sqrt{h}}{16 \pi G}(h^{ij}K - K^{ij})~.
\end{equation}
One may now write the Einstein-Hilbert action in a form that makes the Hamiltonian density explicit
\begin{equation}
S_{\text{EH}} = \int d^4 x\, \sqrt{h}\, \left(\pi^{ij} \frac{\partial h_{ij}}{\partial t}-N \mathcal{H}[\pi^{ij}, h_{ij}]-N_i\mathcal{H}^i[\pi^{ij},h_{ij}]\right)~,
\end{equation}
where we have defined
\begin{eqnarray}
    \mathcal{H}&\equiv&\frac{8 \pi G}{\sqrt{h}}G_{ij,kl}\pi^{ij}\pi^{kl}-\frac{\sqrt{h}}{16 \pi G}(R^{(3)}-2\Lambda)~,\\
    \mathcal{H}^i &\equiv& -2\nabla_k \pi^{ki}~.
\end{eqnarray}
%\begin{equation}
%\begin{split}
%\mathcal{H} &= \frac{1}{16 \pi G  \sqrt{h}}(\text{Tr }\boldsymbol\pi^2-\frac{1}{2}(\text{Tr }(\boldsymbol\pi))^2-\sqrt{h}\,(R-2\Lambda))\\
%\mathcal{H}^i &= -\frac{1}{8 \pi G} \nabla_k \pi^{ik}~.
%\end{split}
%\end{equation}
We have further defined the metric on the space of three metrics, 
\begin{equation}
G_{ij,kl}\equiv h_{ik}h_{jl}+h_{il}h_{jk}-h_{ij}h_{kl}~.
\end{equation}
Interestingly, and importantly, $G_{ij,kl}$ has indefinite signature at each spatial point. For instance, the configurations $a(t) \delta_{ij}$ correspond to a negative direction of $G_{ij,kl}$.

As already noted, the $N$ and $N^i$ appear in the action as Lagrange multipliers enforcing the diffeomorphism constraints of our theory. The constraint associated to $N$ is known as the Hamiltonian constraint 
\begin{equation}\label{Hamiltonian constraint}
    \mathcal{H}=\frac{8 \pi G}{\sqrt{h}}G_{ij,kl}\pi^{ij}\pi^{kl}-\frac{\sqrt{h}}{16 \pi G}(R^{(3)}-2\Lambda)=0~.
\end{equation}
The above is given by the $tt$-component of the Einstein equations. It implies all configurations must have vanishing energy.\footnote{This statement persists in the presence of matter fields, where the equation gets modified by adding the $tt$-component of the matter stress-energy tensor. So in a spatially closed universe, galaxies, and stars, and the totality of everything, has vanishing energy.} Given that the potential term in (\ref{Hamiltonian constraint}) $\sqrt{h}R^{(3)}$ carries a negative sign, and $G_{ij,kl}\pi^{ij}\pi^{kl}$ is not positive definite, gravity indeed has the right structure for a large number of configurations with vanishing energy. The gravitational field can also dress positive energy matter field configurations to a vanishing energy configuration.  The momentum constraints stems from those diffeomorphisms tangential to $\Sigma$ and reads
\begin{equation}\label{momentum constraints}
    \mathcal{H}^i=\nabla_k \pi^{ik}=0~.
\end{equation}
The momentum constraints are associated with the shift function, and $ \mathcal{H}^i=0$  corresponds to the time-space components of the Einstein equations.

Finally, there are the dynamical Einstein equations which can also be expressed in terms of the phase space variables, now in the Hamilton form for $\dot{h}_{ij}$ and $\dot{\pi}_{ij}$.
%Unlike the Einstein field equations where the constraints and dynamical equations appear all in one master set of equations, the canonical theory splits them up.
\subsection{Quantisation}

Given the formulation of general relativity in terms of phase space variables, one can propose a method of canonical quantisation. This was done in early work of DeWitt \cite{dewitt1967quantum,DeWitt:1967ub}. One promotes the classical field profiles at a fixed time to Hermitean operators, $\hat{h}_{ij}(x^i)$, and $\hat{\pi}_{ij}(x^i)$, satisfying the canonical commutation relations
\begin{equation}
\left[\hat{h}_{ij}(x^i),\hat{\pi}_{kl}(y^i)\right] = i \delta_{ik}\delta_{jl}~\delta^{(3)}(x^i-y^i)~.
\end{equation}
Physical states in the Hilbert space, $\mathcal{H}_{\text{grav}}$ are then declared to be those annihilated by the quantum Hamiltonian and momentum constraints. 

It is particularly useful to consider the problem in a Schr\"odinger formalism (see \cite{Anninos:2014lwa} for a perturbative overview). Here one defines eigenstates, $|{h}_{ij}(x^i)\rangle$ of the field operator $\hat{h}_{ij}(x^i)$, and represents states as functionals of $h_{ij}(x^i)$, much like we would in the Schr\"odinger picture of ordinary quantum field theory. In particular, of a state $|\Psi\rangle$ we have the wavefunctional $\Psi[h_{ij}(x^i)] = \langle h_{ij}(x^i)| \Psi\rangle$. In principle,  $\Psi[h_{ij}(x^i)]$ should also depend on the time coordinate, $t$, however the Hamiltonian constraint informs us that there cannot be such a dependence. This is a result of the fact that time reparametrisations are redundancies in our theory. In the field basis representation, we also have that $\hat{\pi}_{ij}=-i \frac{\delta}{\delta h^{ij}}$.

The quantum versions of the Hamiltonian constraint (\ref{Hamiltonian constraint}) and momentum constraints (\ref{momentum constraints}) are respectively referred to as the Wheeler-DeWitt and momentum equations. They are expressed as functional  equations acting on the wavefunctionals $\Psi[h_{ij}(x^i)]$. Explicitly, they are given by
\begin{eqnarray}\label{WdW}
    \left(-\frac{8\pi G}{\sqrt{h}} \,G_{ij,kl} \frac{\delta}{\delta h_{ij}} \frac{\delta}{\delta h_{kl}}-\frac{\sqrt{h}}{16 \pi G}\left(R^{(3)} -2\Lambda \right)\right) \Psi[h_{ij}] &=& 0~, \\
     \nabla_i \left(\frac{-i}{\sqrt{h}} \frac{\delta\Psi[h_{ij}]}{\delta h_{ij}}\right) &=& 0~.
\end{eqnarray}
The momentum constraint ensures the wavefunctional is invariant under spatial diffeomorphisms of $\Sigma$. Physically, this implies that $\Psi[h_{ij}(x)]$ depends only on invariant properties of the three-geometry. The Hamiltonian constraint is an expression of the lack of time evolution at the quantum level.
\newline\newline
{\textbf{Wheeler-DeWitt Inner Product?}} Even if we are able to solve the Wheeler-DeWitt equation, and find a rich class of solutions $\Psi[h_{ij}]$, we are still left with the question of how to formulate the inner-product structure of the gravitational Hilbert space. This is crucial, if we are to extract meaningful expectation values from the physical states in $\mathcal{H}_{\text{grav}}$. One might expect that such an inner product should render Hermitian the operators $\mathcal{H}$ and $\mathcal{H}_i$, as they are real in the classical theory. However, one recalls at this stage that `kinetic term'  $G_{ij,kl}\pi^{ij}\pi^{kl}$ is not positive definite. Rather, the Wheeler-DeWitt equation has the structure of a functional version of the Klein-Gordon equation. As such, it has been proposed from the early days of Wheeler-DeWitt theory that one should consider the Klein-Gordon inner-product. It has the virtue of being conserved but the drawback of not being positive definite. Therefore the Klein-Gordon norm lacks a direct probabilistic interpretation. What this might mean is that we should take a `constant time' slice, $\mathcal{T}$, in the space of three-metrics along which the induced part of $G_{ij,kl}\pi^{ij}\pi^{kl}$ is positive definite. If we denote elements of $\mathcal{T}$ as $\tilde{h}_{ij}$, schematically, the Klein-Gordon type inner product for the Wheeler-DeWitt equation reads
\begin{equation}
(\Psi_1,\Psi_2)_{\text{KG}} = \frac{i}{\text{vol} \, \mathcal{G}_{\text{diff}}}\int_{\mathcal{T}} [\mathcal{D}{\tilde{h}_{ij}}]  \left( \Psi_1^*[\tilde{h}_{ij}] {\nabla_{\tau_{ij}(x^i)} }  \Psi_2[\tilde{h}_{ij}] - \Psi_2[\tilde{h}_{ij}]{\nabla_{\tau_{ij}(x^i)} }  \Psi_1^*[\tilde{h}_{ij}] \right)~,
\end{equation}
where $\nabla_{\tau_{ij}(x^i)}$ represents a covariant variational derivative in the direction normal to the hypersurface $\mathcal{T}$. The Klein-Gordon inner product, $(\Psi_1,\Psi_2)_{\text{KG}}$, must be further gauge fixed to be made sense of since it will naively diverge due to gauge redundancies. Conceptually, it has the virtue of being preserved with respect to the normal direction to $\mathcal{T}$, which might be viewed as a time variable built into the Wheeler-DeWitt formalism. However, $(\Psi_1,\Psi_2)_{\text{KG}}$ is not positive definite and as such difficult to make sense of as an inner product yielding probabilities for physical phenomena without imposing additional conditions. Perhaps such conditions must be part of any underlying theoretical description.\footnote{In the case of spacetimes with asymptotic boundaries, the Wheeler-DeWitt equation and inner product becomes clearer due the presence of an asymptotic clock that is rigid and not subject to the  Hamiltonian constraint. Here, we can evolve states with respect to the ADM Hamiltonian. A canonical treatment also reveals that the condition of an asymptotic boundary improves the positivity properties of the inner product $(\Psi_1,\Psi_2)_{\text{KG}}$, as discussed in \cite{Witten:2022xxp}. In the next section, we will discuss the question of general relativity in the presence of boundaries in more detail.}

An alternative inner-product that is sometimes discussed is to simply declare that
\begin{equation}\label{naive norm}
    (\Psi_1,\Psi_2) \overset{?}{=} \frac{1}{\text{vol} \,\mathcal{G}_{\text{diff}}} \int [\mathcal{D}h_{ij}]\, \Psi_1^*(h_{ij}) \Psi_2(h_{ij})~.
\end{equation}
It has the virtue of being positive definite. In this norm, we simply forget about any notion of time. However, in doing so we will be lead to additional divergences. The norm is somewhat analogous to integrating the absolute value squared of a quantum mechanical wavefunction over all configuration space, as well as all time.
% $a$\footnote{Where the scale factor $a$ is introduced in the minisuperspace approximation via $h_{ij}dx^idx^j=a^2d\Omega^2_3$.} ($\sim$ cosmological time). 
\newline\newline
{\textbf{Further canonical challenges.}} The Wheeler-DeWitt equations are difficult to make complete mathematical sense of. For one, they suffer from operator ordering ambiguities. Moreover, the algebra of constraints satisfied by $\mathcal{H}$ and $\mathcal{H}_i$ does not immediately close upon quantisation \cite{henneaux1983poisson,teitelboim1973commutators}, in the least due to the need for regularisation.  From the perspective of effective field theory, we also note that additional corrections to (\ref{WdW}) will appear due to the presence of higher curvature terms in our theory. Further to all this, it is notoriously difficult to solve the Wheeler-DeWitt equations.   On a more conceptual note, $\Psi[h_{ij}]$ is manifestly independent of time. Instead any notion of time must somehow emerge, at least in semiclassical limit, due to some piece of the three-geometry being naturally associated with the passage of time. The reader can find an excellent discussion on this point in section VIII of \cite{Weinberg:1989}.

\subsection{Some features of Wheeler-DeWitt solutions}

We have already noted that the Wheeler-DeWitt equation is notoriously hard to solve. However, there is one piece of the equation that admits a more tractable treatment, namely the large volume limit for $h_{ij}$ for a theory with $\Lambda>0$. Let us parametrise this limit by taking $h_{ij} = \upsilon^2 \,\tilde{h}_{ij}$, and working in the large-$\upsilon$ limit. To leading order in the large volume limit, the Wheeler-DeWitt equation (\ref{WdW}) simplifies significantly. We refer to \cite{Pimentel:2013gza-ch04} for a detailed treatment. What happens in this limit, is that the potential term in $\mathcal{H}$ is driven to large values, effectively making this a type of functional WKB problem. As such, we can parametrise our wavefunctionals in the semiclassical form $\Psi[h_{ij}] \approx \exp i {W}[h_{ij}]$, where $W[h_{ij}]\in\mathbb{C}$ is some functional of the three-metric. To leading order in the semiclassical limit, we find that the Hamiltonian constraint reduces to
\begin{equation}
\frac{8 \pi G}{\sqrt{h}}G_{ij,kl}\frac{\delta W}{\delta h_{ij}}\frac{\delta W}{\delta h_{kl}}+\frac{1}{16 \pi G}\sqrt{h} \left(-R^{(3)}+2\Lambda\right)=0~.
  %  \frac{8 \pi G}{\upsilon^3 \sqrt{h}}G_{ij,kl}\frac{\delta W}{\delta h_{ij}}\frac{\delta W}{\delta h_{kl}}+\frac{1}{16 \pi G}\sqrt{h} \upsilon^3\left(-\frac{R^{(3)}}{\upsilon^2}+2\Lambda\right)=0~,
\end{equation}
%where we have kept $\upsilon$ as a bookkeeping device for the large volume limit. 
%The above equation is significantly simpler than the original Hamiltonian constraint equation. 
In a large-$\upsilon$ expansion, the above equation is  solved by
\begin{equation}
    W[h_{ij}] = \alpha_0 \int d^3x \sqrt{h}+ \alpha_1 \int d^3x \sqrt{h}R^{(3)}+W_0[h_{ij}] + \ldots~,
\end{equation}
with $\alpha_0 = \frac{1}{4\pi G}\sqrt{\frac{\Lambda}{3}}$, and $\alpha_1 = - \frac{1}{16\pi G}\sqrt{\frac{3}{\Lambda}}$. The functional $W_0[h_{ij}]$ satisfies the variational equation
\begin{equation}\label{conf}
    h_{ij} \frac{\delta}{\delta h_{ij}} W_0[h_{ij}]=0~.
\end{equation}
We thus find that in the large volume regime, the wavefunctionals become rapidly oscillating, with a phase that is expressed locally in terms of $h_{ij}$. The order $\upsilon^0$ term, namely (\ref{conf}), imposes that $W_0[h_{ij}]$ is a Weyl invariant functional of $h_{ij}$. For a four-dimensional theory, this implies that $W_0[h_{ij}]$ is generally a non-local functional of $h_{ij}$. Thus, $W_0[h_{ij}]$ encodes the non-local correlations of the metric field in the given state. The Weyl-invariance of the absolute value $|\Psi[h_{ij}(x^i)]|=|\Psi[e^{2\omega(x^i)}h_{ij}(x^i)]|$ is the quantum analogue of the conformal structure near $\mathcal{I}^+$ of asymptotically de Sitter spacetimes (\ref{FG}) at the classical level. 

Returning back to the naive norm, (\ref{naive norm}), we see that the divergences it produces in the large volume limit correspond to the volume of $\text{diff}\times\text{Weyl}$ gauge group. Formally, to render it finite we would have to divide by the corresponding volume of this redundancy, yielding
\begin{equation}
      (\Psi_1,\Psi_2) \overset{?}{=}  \frac{1}{\text{vol} \, \mathcal{G}_{\text{diff}\times\text{Weyl}}} \int [\mathcal{D}h_{ij}] \,\Psi_1^*[h_{ij}] \Psi_2[h_{ij}]~.
\end{equation}
However, to make the above expression meaningful we still require a careful BRST treatment to fix the infinite gauge redundancies. At least perturbatively close to the expanding part of a pure de Sitter configuration, this inner product appears to yield sensible results. We emphasise, however, that even perturbatively the above treatment requires treating the de Sitter isometries as pure gauge transformations \cite{Higuchi:1991tm}. Recent developments along these lines include \cite{Anninos:2017eib,Chakraborty:2023yed,Collier:2025lux,Cotler:2025gui}.
\newline\newline
{\textbf{Hartle and Hawking's solution.}} In \cite{hartle1983wave} Hartle and Hawking proposed to construct a Wheeler-DeWitt wavefunction via the Euclidean gravitational path integral. Their proposal was motivated by drawing an analogy to the construction of ground states in quantum mechanics from the perspective of the path integral. Particularly, one does so by taking the configurations to live along a complexified time coordinate and imposing a condition at the infinite Euclidean past. 

What Hartle and Hawking proposed for the gravitational case, is to perform a gravitational path integral weighted by the Euclidean gravitational action $S_E[g_{\mu\nu}]$, supplemented with appropriate boundary terms, over a four-manifold, $\mathcal{M}$, with a single boundary, $\partial\mathcal{M}$. The boundary $\partial\mathcal{M}$ of $\mathcal{M}$ is viewed as a spatial Cauchy surface along which measurements are made. The absence of any other boundaries in $\mathcal{M}$ is meant to encapsulate the gravitational analogue of the ground state condition for quantum mechanical path integrals. The Hartle-Hawking path integral is thus given by
\begin{equation}\label{HH state}
    \Psi_{\text{HH}}[h_{ij}] = \sum_{\mathcal{M}} \frac{1}{\text{vol}\,\mathcal{G}_{\text{diff}}}\int [\mathcal{D}g_{\mu\nu}] e^{-S_E[g_{\mu\nu}]}~,
\end{equation}
where $\mathcal{M}$ is a manifold with a single boundary, and we have allowed for the possibility of summing over  multiple single boundary topologies. At the boundary $\partial \mathcal{M}$, we must fix some data. To comply with the standard form of the Wheeler-DeWitt equation  this data is often taken to be the induced metric $h_{ij}$ on $\partial \mathcal{M}$. An alternative choice, discussed in the next section, is to fix the trace of the extrinsic curvature, $K$, along with the conformal class of the metric at $\partial\mathcal{M}$. The latter case is sometimes known as the York time parametrisation \cite{York:1972sj}. Though we will not do so, it is straightforward to incorporate matter fields into the Hartle-Hawking construction. 
\newline\newline
{\textbf{Hartle-Hawking challenges.}} To make sense of the Euclidean path integral (\ref{HH state}) one must deal with the conformal mode problem \cite{gibbons1978path}. This problem stems from the fact that, unlike ordinary quantum field theories in Euclidean signature, the Euclidean action of general relativity is unbounded from below. As such, the path integral over the real contour of Euclidean metrics diverges badly, and one needs to understand how to modify it to render the Euclidean path integral sensible. This is typically done by employing a saddle point approximation, and judiciously selecting physically reasonable saddles. But there is no complete mathematical framework for how to treat the Euclidean path integral.\footnote{Significant evidence in its favour stems from the AdS/CFT correspondence in Euclidean signature, where we have a good notion of Euclidean conformal field theory that is expected to be dual to a Euclidean theory of gravity with anti-de Sitter asymptotia. Being a local issue, the conformal mode problem is present irrespective of the sign of $\Lambda$.} In practice, computing (\ref{HH state}) in generality is hard. In the semiclassical approximation one can study perturbative features around a semiclassical saddle. Even so, one still encounters a linearised version of the conformal mode problem that must be treated \cite{gibbons1978path,polchinski1989phase}. Alternatively, the minisuperspace approach turns off all but a finite number of degrees of freedom in $h_{ij}(x^i)$, and hopes to derive some intuition about the non-linear features of the wavefunction. It is important to emphasise, however, that the minisuperspace approach is not a systematically controllable approximation scheme. Rather, it is a coarse truncation of the underlying problem.

\section{Gravity in finite regions}

In this section, we study general relativity in a finite region. One motivation stems from the fact that in cosmological spaces we generally have no asymptotic spatial or null boundaries where we can anchor physical observables. In such circumstances, diffeomorphism invariance can lead to complications, since truly coordinate invariant quantities are hard to come across and are rarely, if ever, mathematically described in a local fashion.\footnote{This is not to be confused with the fact that physical phenomena occur over arbitrarily small (or vast) regions of spacetime!} To amend this, one can consider some type of relational observable, where we associate the physical behaviour of a system to that of another. For instance, we might measure the passage of time in cosmology through the average temperature of the Universe.

\begin{figure}[t]
	\centering
	\includegraphics{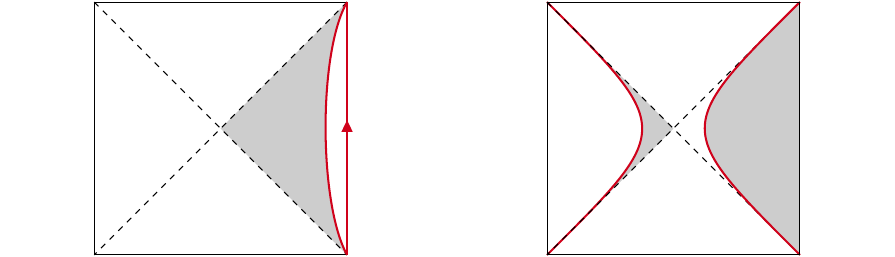}
        \caption{Two examples of gravity in finite regions relevant for cosmology. On the left, the red arrow represents the worldline of an observer in de Sitter space. We can thicken this worldline in a worldtube, and study general relativity in the shaded region, whose boundary is the boundary of the worldtube. On the right, the red lines represent the stretched horizon, a timelike surface close to the cosmological horizon. We can study general relativity on either side of the stretched horizon. }
	\label{fig:worldline_stretched_horizon}
\end{figure}

When we make measurements in our own Universe, these are often performed  quasi-locally.\footnote{To get a sense of scales, we can estimate the ratio between the size of the largest telescope to the size of the observable universe. This comes out to be around $10^{-24}$. We can compare this to the estimate of the ratio of the size of the ATLAS detector to the 10TeV distance scale being probed at CERN, which turns out to be around $10^{-21}$. We see that the cosmological measurements that we perform are taking place on a very small region of spacetime indeed.} We collect properties of distant light in our terrestrial observatories, and then process this into a detailed picture of the large scale structure of the cosmos. One way to model this situation is by introducing an idealised observer moving along a given worldline in a cosmological spacetime, or a slightly thickened worldline that can be viewed as the worldtube of a gravitational observatory. To capture the observatory's experience we consider studying gravity on a manifold in the presence of a finite size worldtube boundary. We depict this in Figure~\ref{fig:worldline_stretched_horizon}. As a final remark, we also mention that another type of finite size timelike boundary appears in the literature is  the stretched horizon --- a timelike hypersurface located close to the de Sitter cosmological event horizon. 

To recapitulate, a more quasi-local understanding of general relativity may shed light on the problem of gravitational observables and nature of cosmological and/or black hole horizons.

\subsection{General relativity with a timelike boundary}

To make the problem more concrete, we would like to discuss general relativity on a manifold $\mathcal{M}$ with a timelike boundary $\Gamma=\partial\mathcal{M}$, such that the spatial section of $\Gamma$ (the intersection of $\Gamma$ with a Cauchy slice $\Sigma$) has finite proper volume. This section is mostly based on \cite{anninos2025gravitationalobservatories,anninos2025cosmologicalobservatories} and references therein. The setup should be contrasted with the asymptotic boundaries that are usually considered, for example in the AdS/CFT correspondence.  In the AdS/CFT correspondence, a gravitational theory in anti-de Sitter space is dual to a conformal field theory `living' at the asymptotic timelike boundary of the spacetime. The absence of an asymptotic timelike boundary in cosmology makes the application of this paradigm difficult. 

For example, global de Sitter space has two asymptotic boundaries, $\mathcal{I}^+$ and $\mathcal{I}^-$, and these present several difficulties. First, unlike in the case of anti-de Sitter space, $\mathcal{I}^\pm$ are spacelike boundaries, concealing the explicit details of bulk time evolution. Second, only a finite portion of $\mathcal{I}^\pm$ is accessible to any given observer.  Thus, the physical relevance of $\mathcal{I}^\pm$ for describing the experience of an observer in an expanding universe is suspect. Finally, unlike the case of anti-de Sitter space, metric perturbations in the bulk do not die off at $\mathcal{I}^+$, and therefore the AdS/CFT procedure of fixing an asymptotic metric on the boundary is inapplicable. 

Given these difficulties, it appears well motivated to introduce a timelike boundary, either as an auxiliary device or otherwise, and consider gravity in the ensuing finite region. The two aforementioned cases, the worldline/worldtube perspective \cite{Anninos_2012,Coleman_2022,Banihashemi_2022,witten2023algebrasregionsobservers,loganayagam2024influencephasedsobserver,Narovlansky:2023lfz} and the stretched horizon \cite{Banks2003,BanksFiolMorisse2006,AnninosEtAl2012,Susskind2021,AnninosHarris2021,Shaghoulian2022a,ShaghoulianSusskind2022,BanihashemiEtAl2023}, are being actively explored in the context of de Sitter holography. 

An important element of general relativity with finite boundaries is the determination of a reasonable boundary condition. In the mathematics literature, what one is after is a notion of  well posedness,  such that the evolution equations allow for the existence of solutions developing uniquely from initial data in a sufficiently smooth manner. We first review, briefly, the initial value problem of general relativity. We then present three different choices of boundary conditions for general relativity with a finite size timelike boundary. 
%We will argue that the most common Neumann and Dirichlet boundary conditions are generically not well posed. A different set of boundary conditions, called conformal boundary conditions, appears to give some promising results.

\subsection{Initial value \& initial boundary value problems}

The initial value problem of classical general relativity is the following \cite{Foures-Bruhat1952,Arnowitt_2008,Stachel1969}. We consider a manifold $\mathcal{M}$ with a Cauchy slice $\Sigma$, and we fix the induced metric $h_{ij}$ and the extrinsic curvature $K_{ij}$ along $\Sigma$. We refer to $h_{ij}$ and $K_{ij}$ as the Cauchy data along $\Sigma$. We are tasked with finding a metric $g_{\mu\nu}$ that satisfies the Einstein field equations on $\mathcal{M}$ and has the correct initial values on $\Sigma$,
\begin{equation}
    e^{\mu}_i e^{\nu}_j g_{\mu\nu} |_\Sigma =h_{ij}~,
    \qquad 
    e^{\mu}_i e^{\nu}_j \nabla_{\mu} n_\nu |_\Sigma = K_{ij}~.
\end{equation}
Here $e^{\mu}_i$ are tangent vectors to $\Sigma$, and $n^\mu$ is a normal vector to $\Sigma$. The projections
\begin{equation}
   n^\mu n^\nu G_{\mu\nu}=0~,
   \qquad 
   n^{\mu} G_{\mu\nu}=0~,
\end{equation}
of the Einstein equations in the  direction normal to the hypersurface $\Sigma$ contain only first derivatives with respect to time, and therefore act as constraints on the initial data. The constraints on the Cauchy data $\{h_{ij}, K_{ij}\}$, which we already saw in the previous section, read
\begin{equation}
    R^{(3)} + K^2 - K_{ij}K^{ij} - 2\Lambda |_\Sigma = 0~, 
    \qquad 
    \nabla_j(K^{ij}-h^{ij}K)|_\Sigma=0~,
    \label{eq:gravconstr}
\end{equation}
where $R^{(3)}$ is the Ricci scalar associated with the induced metric $h_{ij}$, and $K=h^{ij}K_{ij}$ is the trace of the extrinsic curvature. Any choice of initial data must satisfy these constraints. Furthermore, when these constraints are satisfied, one can show that a solution to the initial value problem exists and is unique. In this case, we say that the initial value problem is well posed. This was established in Minkowski space by pioneering work of Choquet-Bruhat, and then Choquet-Bruhat and Geroch \cite{foures1952theoreme,choquet1969global}. Stability of Minkowski space was subsequently established by Christodoulou and Kleinerman in \cite{christodoulou1993global}.

In a gravitational theory on a manifold with a boundary, $\Gamma$, we must also specify boundary conditions for the metric. This is the initial boundary value problem \cite{Friedrich:1998xt}. The solution to the initial value problem is then further constrained by these additional data. We  consider a manifold, $\mathcal{M}$, with a timelike boundary, $\Gamma$, and assume that $\Gamma$ is finite in the sense that the spatial section of $\Gamma$ has finite proper volume. In the presence of a boundary, the initial value problem must be supplemented by boundary conditions on $\Gamma$ that the solution $g_{\mu\nu}$ must satisfy. We consider three different boundary conditions:
\begin{itemize}
    \item[i.] Dirichlet: fix the induced metric on $\Gamma$~,
    \item[ii.] Neumann: fix the extrinsic curvature of $\Gamma$~,
    \item[iii.] Conformal: fix the conformal class of the induced metric and the trace of the extrinsic curvature on $\Gamma$~.
\end{itemize}
In addition to the conditions on $\Gamma$, one must also specify Cauchy data at some initial time surface $\Sigma$ that intersects $\Gamma$, as well as appropriate compatibility conditions, the corner conditions, at the intersection of $\Sigma$ with $\Gamma$.  

\subsection{Dirichlet/Neumann problem for a finite boundary}

The Dirichlet and Neumann problems are natural to consider and have appeared in many contexts throughout the physics literature. It is interesting to note that they do not generically give rise to a well posed initial value problem. In particular, they lack existence and uniqueness properties for the most generic case, both in Lorentzian \cite{An_2021,Anninos:2022ujl} and Euclidean signature \cite{Avramidi:1997sh,Anderson_2008,witten2023noteboundaryconditionseuclidean}.  
\newline\newline
{\textbf{Lorentzian Dirichlet problem.}} Consider Dirichlet boundary conditions, where we fix the induced metric $g_{mn}$ on a timelike boundary $\Gamma$. Suppose that the metric  $g_{mn}$ satisfies the constraint equation along $\Gamma$
\begin{equation}
   \left( R^{(3)} + K^2 - K_{mn}K^{mn} - 2\Lambda \right) |_\Gamma~.
\end{equation}
With Dirichlet boundary conditions, we are free to choose a slightly different metric, $g_{mn} + \delta g_{mn}$ along $\Gamma$, for generic $\delta g_{mn}$. The linearised constraints read
\begin{equation}
 \left(   \delta R^{(3)} - 2K \, \delta K 
    + 2 K^{mn} \delta K_{mn}
    + 2 K^{mr} K_{\phantom{n}r}^{n} \delta g_{mn} \right)|_\Gamma = 0~.
\end{equation}
At least near sufficiently flat regions,  this equation does not hold for generic choices of $\delta g_{mn}$ and therefore contradicts the assumption that we can freely fix the metric on $\Gamma$  \cite{Anderson_2008,An_2021}. 
%In other words, one cannot satisfy the constraint equations \eqref{eq:gravconstr} for arbitrary choices of $g_{mn}$. 
Thus, it is said that Dirichlet boundary conditions are not well posed for finite boundaries. We emphasise, however, that this does not mean that there cannot be special choices of Dirichlet boundary conditions that yield improved results. Indeed, there is evidence for certain cases of a good Dirichlet problem for small variations of initial data for spherical or near spherical spatial boundaries \cite{anninos2025gravitationalobservatories}. A similar discussion  holds for Neumann boundary conditions.
\newline\newline
{\textbf{Euclidean Dirichlet problem.}} For general relativity in Euclidean signature on a manifold with a finite size boundary, it has been shown in  \cite{Avramidi:1997sh,Anderson_2008,witten2023noteboundaryconditionseuclidean} that Dirichlet boundary conditions do not give rise to an elliptic problem. As such, the Dirichlet problem cannot be used, generically, as a basis for perturbation theory. An elliptic operator $\mathfrak{G}$ is a differential operator which has at most a finite set of zero modes, and can be inverted. On a manifold with boundary, the boundary conditions select the allowed zero modes and determine whether an operator $\mathfrak{G}$ is elliptic.  For the Euclidean Dirichlet problem close to a flat region, one finds an infinite number of zero modes corresponding to bulk diffeomorphisms that cannot be removed by a gauge-fixing procedure defined purely on the boundary.  Again, despite all this, there can be special choices of Dirichlet data that yield improved conditions \cite{witten2023noteboundaryconditionseuclidean}. A similar discussion holds for Neumann boundary conditions. %Ellipticity is a necessary property in order to define a quantum perturbation theory, since the Green function is obtained by inverting the kinetic operator. We conclude that Euclidean gravity with Dirichlet boundary conditions does not admit a sensible perturbation theory.
\newline\newline
{\textbf{The case of an asymptotic boundary.}} Let us contrast what we have seen so far with the case of asymptotic boundaries. Consider, for instance, an asymptotically anti-de Sitter boundary. In AdS$_4$/CFT$_3$, the conformal class of the induced metric at the boundary acts as the source for the stress tensor in the dual conformal field theory. Due to the asymptotic nature of the boundary, the arguments leading to non-existence, non-uniqueness, or non-ellipticity no longer apply. Both Einstein equations and the ensuing solution space changes qualitatively near the anti-de Sitter boundary. Near the asymptotic boundary, we can expand the metric in a large $\rho$ expansion \cite{fefferman1985conformal} as
\begin{equation}\label{fgads}
     ds^2 = d\rho^2 + \left(e^{2\rho} g_{mn}^{(0)} + g_{mn}^{(2)} + e^{-\rho} g_{mn}^{(3)} + \dots \right) dx^m dx^n~,
\end{equation}
where we have set the cosmological constant $\Lambda=-3$, or equivalently the anti-de Sitter length $\ell=1$. The constraint equations are satisfied at leading order in $\rho$ for any choice of the induced metric $g_{mn}^{(0)}$. The extrinsic curvature near the boundary behaves as $K_{mn} \approx e^{2\rho} g_{mn}$. The trace of the extrinsic curvature, $K$, evaluated at the boundary is constant and finite, namely $K=3$, irrespective of the choice  $g_{mn}^{(0)}$. Moreover, one should view the boundary metric $g_{mn}^{(0)}$ as a conformal metric rather than being totally fixed. Indeed, there are Fefferman-Graham preserving diffeomorphisms that map (\ref{fgads}) to itself under an appropriate conformal mapping of the boundary data. This is the bulk manifestation of the fact that the dual three dimensional conformal field theory only perceives the conformal structure of the three-metric $g^{(0)}_{mn}$. 

So the boundary conditions of an asymptotically four-dimensional anti-de Sitter space have a conformal flavour. Rather than a Dirichlet problem, we could instead state that we are fixing the conformal class of the boundary metric along with the special condition $K=3$.
%Thus, Dirichlet boundary conditions are well posed for asymptotic boundaries.

\subsection{Conformal boundary conditions}

We now consider boundary conditions we will refer to as conformal boundary conditions. In this case, we fix the induced metric on the boundary, $\Gamma$, only up to a conformal factor alongside the trace of the extrinsic curvature,
\begin{equation}
   e^{\mu}_m e^{\nu}_n g_{\mu\nu} |_\Gamma= \Omega(x^m)\, g_{mn}~,
   \qquad 
   K_{mn} g^{mn} |_\Gamma = K~.
\end{equation}
Here, $g_{mn}(x^m)$ and $K(x^m)$ are fixed, but the conformal factor of the induced metric at $\Gamma$, $\Omega(x^m)$, is not fixed. Any solution $g_{\mu\nu}$ to the Einstein equations such that there exists a function $\Omega(x^m)$ for which the above equations hold is a valid solution to the initial value problem in finite regions with conformal boundary conditions.

In Euclidean    signature,  conformal boundary conditions were proven to be elliptic in \cite{Anderson_2008}. In Lorentzian signature, the conformal boundary conditions have been conjectured to be well posed \cite{An_2021}, with some support at the linearised level around a flat boundary \cite{An:2025rlw,Anninos:2022ujl}. However, the general problem remains significantly open-ended. For curved boundaries there are strongly growing modes that appear in the linearised analysis \cite{anninos2025gravitationalobservatories,Liu:2025xij} whose physical meaning remains to be understood, and they appear to disrupt a standard notion of well posedness. Intriguingly, the growing modes are better behaved when the timelike boundary approaches the stretched horizon limit \cite{Anninos:2025zgr}. 
%Conformal boundary conditions and finite boundaries are studied in \cite{Anninos:2023epi} for vanishing cosmological constant, while the analysis is extended to positive cosmological constant in \cite{anninos2025cosmologicalobservatories}. 
\newline\newline
{\textbf{Cosmological observatories.}} Let us briefly summarise some of the results of \cite{anninos2025cosmologicalobservatories,Anninos:2025zgr}. Here a timelike boundary is placed in a $\Lambda>0$ theory, and subjected to conformal boundary conditions. A finite boundary is introduced at some fixed radial value, $r$, in the static patch coordinate system (\ref{static_dS_metric}). The trace of the extrinsic curvature, $K$, is taken to be constant along this hypersurface.  We discuss here the case where its ambient space contains the cosmological event horizon.  The worldline ($K\ell\to-\infty$) and stretched horizon ($K\ell\to\infty$) limits can be obtained by moving the boundary close to the poles or to the cosmological horizon, respectively.

In Euclidean signature, invoking  the methods  employed for Euclidean black hole thermodynamics to the case at hand, provided the finite boundary is sufficiently close to the de Sitter horizon it enjoys a positive specific heat. This is to be contrasted with the Dirichlet case \cite{Coleman_2022,Banihashemi_2022}. The specific heat of the de Sitter horizon in the presence of a Dirichlet boundary is found to be negative.

In Lorentzian signature one can study gravitational perturbations of a de Sitter metric subject to conformal boundary conditions. In the strict stretched horizon limit, corresponding to $K\ell \to \infty$, the majority of modes decay at late times \cite{anninos2025cosmologicalobservatories,Anninos:2025zgr}. A subset of modes organise themselves into fluid dynamical shear and sound modes. These are interpreted as the degrees of freedom of a conformal fluid living near the stretched horizon \cite{AnninosEtAl2012,Anninos:2025zgr}. The  shear viscosity to entropy ratio of this fluid is given by $\tfrac{1}{4\pi}$ which also appears, famously, in other settings \cite{Damour:1979wya,Price:1986yy,policastro2001shear}. Another subset exhibits quasinormal mode behaviour. Finally, there is a set of modes, for each angular momentum sector, that has an imaginary frequency $\omega_\pm = \pm 2\pi i T_{\text{dS}}$, where $\omega$ is the frequency measured in units of the static patch time. The de Sitter horizon temperature, $T_{\text{dS}} = (2\pi\ell)^{-1}$.

A complete physical understanding of these phenomena may, potentially, lead to a finite portion of  de Sitter space with a timelike boundary that is under thermodynamic and dynamical control, at least at the linearised level. 
\newpage

\section{Some homework problems}

\begin{itemize}
    \item[] \textbf{Question 1:} Show that the de Sitter spacetime has a cosmological horizon, and compute its size. \\ 
    % \textbf{A:} The cosmological horizon is the maximum physical distance that can be covered by a signal sent at some time t.    The de Sitter metric is given by 
    % \begin{align}
    %     ds^2=-dt^2+a^2(t)dx^idx^i
    % \end{align}
    % and where $a(t)=e^{t/l}$ for light $ds^2=0$
    % \begin{align}
    %     dx=\frac{dt}{a(t)}
    % \end{align}
    % This is the comoving distance travelled by a signal and the physical distance is given by multiplying with expansion parameter $a(t)$
    % \begin{align}
    %     d_\text{physical}=e^{t/l}\int_t^\infty dt'\;e^{-t'/l}=e^{t/l}(-l)(e^{-t'/l})\Bigg|_t^{\infty}=l
    % \end{align}
    % The cosmological horizon of dS is the dS radius itself.
    %
    \item[] \textbf{Question 2:} Compute the $\nu$ in $\Delta=\frac{3}{2}+i\nu$ for Standard Model neutrino and electron, given the observed value of $\Lambda$. Deduce what $SO(1,4)$ UIRs they furnish.\\ 
    % \textbf{Ans:} We use eq.(\ref{conf_high_spin}) for $s=\frac{1}{2}$ to get
    % \begin{align}
    %     \nu=ml
    % \end{align}
    % To get mass of particles in $\text{length}^{-1}$ units we use expression for Compton wavelength,
    % \begin{align}
    %     \lambda=\frac{h}{mc}
    % \end{align}
    % and substitute for the required quantities to get
    % \begin{align}
    %     m_e\sim10^{12}\;m^{-1} \qquad\qquad m_\nu\sim 10^4\;m^{-1}\\\nn 
    %     \Rightarrow \nu_e=m_e\times l=10^{12}\times10^{26}=10^{38}\\\nn  \nu_\nu=m_\nu\times l=10^4\times 10^{26}=10^{30}
    % \end{align}
    % Such high values of $\nu$ shows that elementary particles of the Standard Model are very heavy with respect to the de Sitter radius.
    %
    \item[] \textbf{Question 3:} In two-dimensional de Sitter space, consider the eigenstates $\ket{n}$ of the generator $L_0$, with unit norm, that obey $L_0\ket{n}=n\ket{n}$ with $n\in\mathbb{Z}$. Compute the norm of the state $L_+\ket{n}$. \\ 
    % \textbf{Ans:} The norm of $\ket{\Psi}$ can be written in terms of ladder operators 
    % \begin{align}
    %     \braket{\Psi}=\bra{n}L_+^{\dagger}L_+\ket{n}=\bra{n}L_-L_+\ket{n}
    % \end{align}
    % where we have used the fact that $L_+^\dagger=L_-$. We use the commutation relation $[L_-,L_+]=-2L_0$ to get the following relation
    % \begin{align}\label{expectation of commutator}
    %     \bra{n}L_-L_+\ket{n}=\bra{n}L_+L_-\ket{n}-2\bra{n}L_0\ket{n}=\bra{n}L_+L_-\ket{n}+2n
    % \end{align}
    % We can further simplify by calculating the expectation value of the quadratic Casimir operator.
    % \begin{align}
    %     \bra{n}C_2\ket{n}&=\bra{n}\l(\frac{L_+L_-+L_-L_+}{2}-L_0^2\r)\ket{n}=\lambda(1-\lambda)\\\nn 
    %     &=\l(\frac{\bra{n}L_+L_-\ket{n}+\bra{n}L_-L_+\ket{n}}{2}-n^2\r)=\lambda(1-\lambda)\\\nn 
    %     &=\bra{n}L_-L_+\ket{n}-n-n^2=\lambda(1-\lambda)
    % \end{align}
    % where we have used eq.(\ref{expectation of commutator}). This gives us the norm of $\mathcal{N}$,
    % \begin{align}
    %    \bra{n+1}|\mathcal{N}|^2\ket{n+1}=\bra{n}L_-L_+\ket{n}=\lambda(1-\lambda)+n+n^2
    % \end{align}
    % $\mathcal{N}$ in general can be a complex number but the additional phase can always be absorbed in the definition of the state itself.
    % \begin{align}
    %     L^\dagger\ket{n}=\sqrt{\lambda(1-\lambda)+n(n+1)}\ket{n+1}
    % \end{align}
    \item[] \textbf{Question 4:} Compute the Harish-Chandra character for principal series UIR of $SO(1,2)$.
    \\ 
%         \textbf{Ans:} The Harish--Chandra character is formally defined as
%     \begin{align}
%     \chi_\nu(t) = \mathrm{Tr}\big(e^{tD}\big).
%     \end{align}

%     On the descendant states $|\Delta_\pm;n\rangle$, we have
%     \begin{align}
%     e^{tD}|\Delta_\pm;n\rangle
%     = e^{-t(\Delta_\pm+n)}|\Delta_\pm;n\rangle,
%     \end{align}
%     where
% \[\Delta_\pm = \frac{1}{2} \pm i\nu.\]
% Therefore, the contribution from each tower is a geometric series (for $\mathrm{Re}\,t>0$),
% \begin{align}
% \chi_{\Delta_\pm}(t)
% = \sum_{n=0}^{\infty} e^{-t(\Delta_\pm+n)}
% = \frac{e^{-t\Delta_\pm}}{1 - e^{-t}}.
% \end{align}
% Since the principal series contains both weights $\Delta_+$ and $\Delta_-$, the full character is
% \begin{align}
% \chi_\nu(t)
% &= \chi_{\Delta_+}(t) + \chi_{\Delta_-}(t) \\
% &= \frac{e^{-t\left(\frac{1}{2}+i\nu\right)} + e^{-t\left(\frac{1}{2}-i\nu\right)}}{1 - e^{-t}}.
% \end{align}
 \item[] \textbf{Question 5:} Show that for $h_{ij}(t,x^i) = a(t)^2 \delta_{ij}$ the classical Hamiltonian constraint becomes one of the Friedmann equations.\\ 
  \item[] \textbf{Question 6:} Compute the ground state wavefunction of a harmonic oscillator from a path integral construction.\\
  \item[] \textbf{Question 7:} Find a portion of pure de Sitter space with a timelike boundary that is thermodynamically and dynamically stable at the linearised level, or prove there is no such case. \\
  \item[] \textbf{Question 8:}  Verify all the equations in the main text.

\end{itemize}

\begin{acknowledgements}
	-- The \href{https://indico.sns.it/event/91/}{{\it La Ricotta} Summer School  {\it The Disordered Universe 2025}} received financial support from the ERC grant NOTIMEFORCOSMO (agreement n${}^{\circ}$ 1126304)), by Scuola Normale and by the Instituto Galego de F{\'i}sica das Altas Enerx{\'i}as (IGFAE-USC/Xunta de Galicia) through the CIGUS network and EU cofinancing program.
\end{acknowledgements}

\bibliographystyle{utphys}
\bibliography{2025-DU-dS-biblio}	

\end{document}